\pdfoutput=1
\documentclass[11pt]{article}
\usepackage[letterpaper,margin=1in]{geometry}
\usepackage[T1]{fontenc}
\usepackage{amsmath,amssymb,amsthm,mathtools}
\usepackage{mathpazo}
\usepackage{graphicx,booktabs,array,multirow}
\usepackage[round,authoryear]{natbib}
\usepackage{enumitem}
\usepackage{microtype}
\usepackage{float}
\usepackage{adjustbox}
\usepackage[font=small,labelfont=bf,labelsep=period,justification=centering]{caption}
\usepackage{xcolor}
\definecolor{linkblue}{RGB}{20,50,110}
\usepackage[colorlinks=true,linkcolor=linkblue,citecolor=linkblue,urlcolor=linkblue]{hyperref}
\newcommand{\tabwrap}[1]{\par\vspace{1pt}{\centering\footnotesize\adjustbox{max width=\textwidth}{#1}\par}}
\newcommand{\floatnote}[1]{\par\vspace{5pt}\noindent\begin{minipage}{\textwidth}\footnotesize #1\end{minipage}}

\makeatletter
\newcommand{\runinperiod}[1]{#1.}
\renewcommand\paragraph{\@startsection{paragraph}{4}{\z@}{1.5ex \@plus 0.5ex \@minus 0.2ex}{-0.6em}{\normalfont\normalsize\bfseries\runinperiod}}
\newcommand{\appendixformat}{%
  \def\@seccntformat##1{\@ifundefined{apx@##1}{\csname the##1\endcsname\quad}{\csname apx@##1\endcsname}}%
  \def\apx@section{Appendix~\thesection.\hspace{0.5em}}%
  \counterwithin{table}{section}\renewcommand{\thetable}{\thesection\arabic{table}}%
  \counterwithin{equation}{section}\renewcommand{\theequation}{\thesection\arabic{equation}}}
\makeatother

\theoremstyle{plain}

\newtheorem{proposition}{Proposition}
\newtheorem{lemma}{Lemma}
\theoremstyle{definition}
\newtheorem{definition}{Definition}
\newtheorem{assumption}{Assumption}

\newcommand{\E}{\mathbb{E}}
\newcommand{\Var}{\operatorname{Var}}
\newcommand{\Cov}{\operatorname{Cov}}
\newcommand{\cF}{\mathcal{F}}
\newcommand{\cG}{\mathcal{G}}
\newcommand{\cH}{\mathcal{H}}

\newcommand{\cL}{\mathcal{L}}

\newcommand{\Fp}{\mathcal{F}^{+}}
\newcommand{\Nw}{N_{w}}
\newcommand{\Div}{\mathit{Div}}
\newcommand{\pd}{\mathit{pd}}

\makeatletter
\newcommand{\N}[1]{\@ifundefined{num@#1}{\textbf{??#1??}}{\@nameuse{num@#1}}}
\makeatother
\setlist{noitemsep,topsep=3pt}
\input{numbers}
\newcommand{\papertitle}{Accounting for News}

\hypersetup{pdftitle={Accounting for News},pdfauthor={Ziyao Wang, Svetlozar T. Rachev, Frank J. Fabozzi}}
\begin{document}
\pagenumbering{Alph}
\title{\papertitle\thanks{The authors received no funding for this work. The authors declare no conflicts of interest. The authors used Claude Fable 5.2 (Anthropic) to polish the language of the manuscript and in part of the data processing. The authors take full responsibility for the content of the article.}}
\author{Ziyao Wang\thanks{Department of Mathematics and Statistics, Texas Tech University. Corresponding author: \href{mailto:ziywang@ttu.edu}{ziywang@ttu.edu}.}
\and Svetlozar T. Rachev\thanks{Department of Mathematics and Statistics, Texas Tech University.}
\and Frank J. Fabozzi\thanks{Carey Business School, Johns Hopkins University.}}
\date{November 2026}
\maketitle
\thispagestyle{empty}
\begin{abstract}
\noindent Language-model embeddings of news explain returns but lack economic units. We define a text's meaning as the revision it induces in expected outcomes. An exact identity linking the outcomes divides the price move the text accounts for among fundamentals, later returns, and valuation. In a decade of 8-K filings, the text accounts for \N{tsB}\% of event-day return variance. Within a week, later returns add about a tenth to that part, mainly in smaller firms, and little to the remainder. That part carries \N{gAFy}\% of the comovement between event-day returns and next-year sales growth, including sales from quarters begun after filing.
\end{abstract}



\clearpage
\pagenumbering{arabic}

\section{Introduction}\label{sec:intro}

A language model turns a piece of news into a vector, called an embedding. Such embeddings forecast stock returns \citep{chen2022expected}, and they account for a measurable part of the return on the day the news appears. An embedding, however, is not an economic object. Its coordinates have no units, any rotation of them fits the data equally well, and nothing in it separates news about next year's earnings from news about the rate at which those earnings are discounted. Since \citet{campbell1988dividend}, the literature has asked how much of a price movement is news about cash flows and how much is news about discount rates. The same question has not been asked of the text that accompanies the movement.

This paper provides a way to ask it. We define the meaning of a text by the expectations that it revises. Consider an outcome that is specified by theory and observed later, such as the return on the day the text becomes public, the growth of sales over the following year, or the return over the following week. The anchored concept of the text for that outcome is the revision in the outcome's conditional expectation brought about by reading the text, holding fixed everything known beforehand. A concept is measured in the units of its outcome. It is unique, so that unlike the coordinates of an embedding it cannot be rotated. It depends on the information that the embedding carries and not on how the embedding is parameterized. Because conditional expectations are linear, an accounting identity that links outcomes links their concepts with the same coefficients.

The identity we use is exact. For any measure of fundamentals, such as book equity or sales, the return over an event window equals the growth of the fundamental through the next $H$ days, plus payouts, minus the returns earned over those $H$ days, plus the change in the ratio of price to the fundamental. No log-linear approximation is involved, the identity holds for firms that pay no dividends, and both sides are observable. The identity divides the part of an event-day return that the text accounts for, the text part, into shares that sum to one at every horizon: the share realized in fundamentals; the share offset by later returns, which is negative if later returns extend the move; and the share still in the valuation ratio. Each share is a regression slope built from covariances, so a single share can be negative or exceed one. The same shares can be computed for the remainder of the event-day return and for returns on days without news. A standard model of persistent cash-flow and discount-rate news shows what several kinds of news would look like in these accounts. News about current cash flows is realized at once. News about growth leaves valuation at the rate at which the growth arrives. Discount-rate news is never realized in fundamentals and leaves valuation only as later returns offset it. We use the model to organize the readings of the evidence; the identity itself records what followed a price move, not why.

We estimate these shares on the current reports that U.S.\ firms file with the Securities and Exchange Commission on Form 8-K. The 466,000 filings of 2014 to 2023, each stamped with the time at which the Commission accepted it, give \N{nFD} firm-days. We read the filings with a language model whose training text ends before the sample begins. The text accounts for \N{tsB}\% of the variance of market-adjusted returns on the event day, the first trading day whose close follows the acceptance of the filing. The share is concentrated in the return from the previous close to the open: when a filing is accepted outside trading hours, the text part of the return is largely in the price by the next open.

Reading the filings with a model trained before the sample does not by itself keep later information out of these estimates: tuning parameters, scaling rules, state variables, and the dates at which accounting data become known can each let it in. We therefore report each main result under two fits. The baseline fit estimates the expectations for each calendar quarter on the other quarters of the decade, less the months whose outcome windows overlap it. The chronological fit uses, at every step, only data dated before the quarter it predicts. In the chronological fit the text accounts for \N{tsC}\% of the variance of event returns in 2017--2023, against \N{tsBlate}\% in the baseline fit for the same years.

Our first finding is that the two parts of an event-day move are followed by different returns. Within a week, later returns add about a tenth to the text part, half of it on the first day, and almost nothing to the remainder of the same day's move, while moves on days without news are partly reversed, as in \citet{chan2003stock} and \citet{savor2012stock}. In the model these three patterns are those of a reaction that is too small, of a complete reaction, and of short-lived discount-rate news or an overreaction. The text part's contribution to the one-week extension of the whole move exceeds what its share of the variance would imply by \N{kZAw} standard errors in the baseline fit and by \N{ckZAw} in the chronological fit. The extension is strongest in the smallest third of firms and is no smaller for filings accepted after the close, which the market has a full trading day to absorb; measured by the offset share, it is present with and without an earnings announcement. In 2017--2023, sorting firm-days on the text part alone, with information available at the close of the event day, gives a spread of \N{sTw} basis points in returns over the next week.

Our second finding is that the text part, a small part of the price move, carries a large part of its relation to later sales. It is about one-twentieth of the variance of an event-day move, and it carries \N{gAFy}\% of the comovement between that move and the growth of sales per share over the following year. The timing of financial reports lets us separate sales by when they were earned: we date each fiscal quarter by the first close after its results are released and split the growth of sales exactly across quarters. A month after an earnings announcement, when the quarter just announced is the only one to have entered the fundamental in 98\% of cases, the text part carries \N{gEFm}\% of the comovement between the announcement return and the growth of sales. On other event days the text part shows no positive relation to sales that had been earned but not yet reported. It is related instead to the sales of the quarter in progress and of the fiscal quarters that began after the filing, and the contribution of the quarters that began after it is \N{vNSAZry} standard errors above what the text part's share of the variance would imply (\N{cvNSAZry} in the chronological fit).

Together the two findings give the text part of a price move a distinct economic signature: later returns extend it, and later sales growth matches far more of it than of the remainder. Both patterns are present in the chronological fit. An embedding, then, does more than predict returns. Anchored to outcomes, it marks out a component of a price move whose later course can be accounted for.

The shares and contributions behind these findings are ratios of variances and covariances of conditional expectations. We estimate them out of sample with the debiased estimator of \citet{chernozhukov2018double}, which is insensitive, to first order, to errors in the fitted expectations, and the standard errors allow for repeated firms, for shocks common to firms on the same and nearby days, and for overlapping outcome windows (Section~\ref{sec:estimator}). Our fitted expectations, which we call readouts, are linear in the embedding, so every statement about the text part concerns the component of the text that a linear function captures. In population the shares of that component sum to one, and in simulations the estimator recovers the variance that the component accounts for and its offset shares. Section~\ref{sec:robust} shows how the estimates vary with the class of readout and with the way it is fitted.

\paragraph{Related Literature} A large literature turns text into numbers and relates the numbers to returns, from dictionaries and tone \citep{tetlock2007giving,loughran2011liability} to supervised word models \citep{ke2019predicting} and to embeddings and prompted scores from large language models \citep{chen2022expected,lopezlira2023chatgpt}; \citet{gentzkow2019text} survey the earlier methods. \citet{didisheim2026inefficient} show that much of the content of news is predictable from firm characteristics and that the unpredictable remainder forecasts returns. In that literature the embedding is a regressor and the return is the object of interest. We study the regressor itself and measure the content of a text in the units of a return identity.

The efforts closest in spirit make text interpretable through topics \citep{bybee2023narrative,bybee2024business} or through concepts that people name in advance \citep{koh2020concept}. Unlike topics and named concepts, anchored concepts are measured in units that add up to the return. How much of return variance can be traced to identifiable news is the question of \citet{roll1988r2}, \citet{boudoukh2019information}, and \citet{brogaard2022moves}. Our measure of it for corporate filings is the share of return variance that the text accounts for, and the identity decomposes the part so identified.

Closest in data are studies of the text of corporate disclosures. \citet{siano2025news} and \citet{wu2025extracting} relate the text of earnings press releases to announcement returns with language models, \citet{costello2024uncovering} measure the information in each filing with firm-specific language models, and \citet{meursault2023pead} measure earnings surprises from the text of earnings calls and document the drift that follows them. These papers ask how much of a return the text predicts. We ask what follows the part that the text accounts for. The estimator builds on the econometrics of machine-generated measurements \citep{chernozhukov2018double,williamson2021nonparametric,ludwig2025large,battaglia2024inference}, and the use of a model trained only on earlier text follows the work on look-ahead bias in language models \citep{sarkar2024lookahead,he2025chronologically,kelly2026scaling}.

Return decompositions begin with \citet{campbell1988dividend} and \citet{campbell1991variance}. \citet{vuolteenaho2002drives} finds that cash-flow news dominates the returns of individual firms, and \citet{lochstoer2020drives} find that it dominates the returns of anomaly portfolios; \citet{chen2013drives} measure cash-flow news with analyst forecasts. Recent work derives identities for scale variables other than dividends \citep{cho2024scale}, decomposes returns with observables in place of a vector autoregression \citep{knox2022decomposition}, and examines the limits of the log-linear identity \citep{bendavid2026crimes}. These decompositions condition on a small set of state variables and say how much of a period's return is cash-flow news. Ours conditions on the text and measures both sides of the identity. The finding that price moves are followed by fundamentals connects to the accounting literature on prices leading earnings \citep{kothari1992information,collins1994lack,lundholm2002bringing} and to measures of price informativeness \citep{bai2016have}.

Section~\ref{sec:framework} develops the framework. Section~\ref{sec:data} describes the data and the estimator. Section~\ref{sec:share} reports the share of event-return variance that the text accounts for, and Section~\ref{sec:fateresults} the decomposition of event-day price moves. Section~\ref{sec:conclusion} concludes.

\section{Framework}\label{sec:framework}

This section defines the economic meaning of a text and derives the accounting used in the empirical work. Figure~\ref{fig:framework} summarizes the construction. Proofs are in Appendix~\ref{app:proofs}. The example of Section~\ref{sec:example} is illustrative: it shows what several kinds of news would look like in the accounts, and we do not use it to identify a mechanism in the data.

\begin{figure}[htbp]
\centering\includegraphics[width=\textwidth]{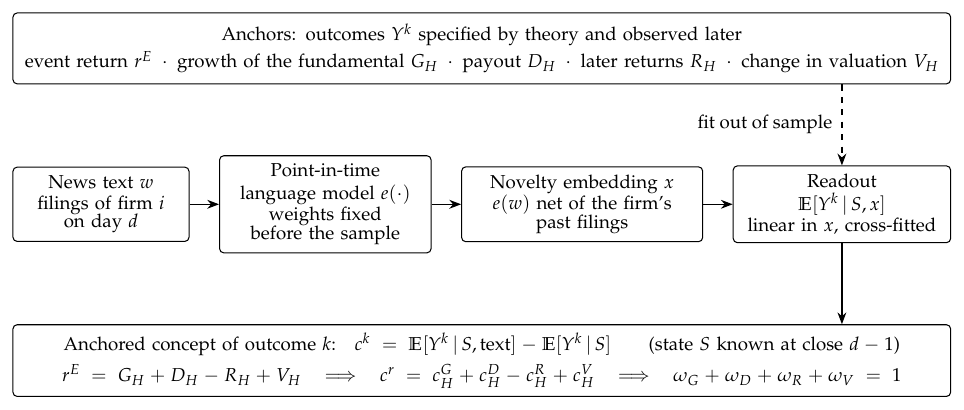}
\caption{Anchoring Text to Outcomes}\label{fig:framework}
\floatnote{News text is encoded by a language model whose weights were fixed before the sample. A readout from the embedding and the state is fitted, out of sample, to outcomes that theory specifies and that are observed after the text. The anchored concept of an outcome is the revision in its conditional expectation due to the text. An exact identity among the outcomes carries over to the concepts, so that the shares of the price move that the text accounts for add to one at every horizon. In the figure, $\omega_{G}$, $\omega_{D}$, $\omega_{R}$, and $\omega_{V}$ are the growth, payout, offset, and valuation shares of Section~\ref{sec:fates}.}
\end{figure}

\subsection{Anchored Concepts}\label{sec:concepts}

Consider a firm and a window of trading time during which a text $w$ about the firm is published; in the application the text is the firm's filings on a given day. Let $\cF$ denote public information before the window and $\Fp$ public information at its end. In the application $\cF$ is generated by a vector $S$ of state variables observed before the window. A reader who knows $\cF$ and reads the text has information $\cG=\cF\vee\sigma(w)$. A language model, the encoder, maps the text to a vector $x=e(w)\in\mathbb{R}^{J}$, its embedding. The results of this section apply equally to any representation that is a function of the text and of information in $\cF$, such as the embedding net of the firm's past filings that we use in the application.

\begin{assumption}[Point in time]\label{ass:pit}
The text $w$ is $\Fp$-measurable, and the encoder $e$ is fixed before the window: it does not depend on information outside $\cF$.
\end{assumption}

Under Assumption~\ref{ass:pit}, $\cF\subseteq\cG\subseteq\Fp$ and $x$ is $\cG$-measurable. The part of any outcome that the text accounts for is then the expectation, given the text, of the news that arrives in the window, and the rest of that news is uncorrelated with it (Lemma~\ref{lem:innov}). The assumption fails when the encoder has been trained on text written after the window, because the embedding may then reflect information that no reader of the text possessed; this is the look-ahead bias studied by \citet{sarkar2024lookahead}. The assumption orders information sets. It does not order the text and the price move inside the window, a point taken up in Section~\ref{sec:share}.

One could try to interpret the embedding itself. The difficulty is that neither the way an embedding is trained nor the way it is used pins down its coordinates. Embeddings are used through a readout: a function, fitted within a class, that maps the embedding and the state to the conditional expectation of an outcome. A class that is closed under invertible linear maps, such as the linear functions, fits $Ax$ exactly as well as $x$ for every invertible matrix $A$, and a training objective that depends on the embedding only through inner products takes the same value at $e$ and at $Ue$ for every orthogonal matrix $U$. The information is in $\sigma(x)$ and not in any coordinate. Methods from the interpretability literature choose coordinates on statistical grounds: some rotate the embedding so that few coordinates are active for any one text, and others find the direction that best predicts a human-assigned label, a linear probe \citep{alain2017understanding,belinkov2022probing}. A label attached in this way does not say how the information in that coordinate is related to prices. We therefore define the meaning of a text by the expectations that it revises.

\begin{definition}[Anchored concept]\label{def:concept}
Let $Y^{k}$ be a square-integrable random variable, the outcome, that is specified by theory and observed at the end of the window or later. The anchored concept of the text for outcome $k$ is $c^{k}=\E[Y^{k}\mid\cG]-\E[Y^{k}\mid\cF]$.
\end{definition}

A concept is the amount by which reading the text revises the expectation of the outcome, and it is measured in the outcome's units: log points of return, or log points of book-equity growth. More generally, for any $\sigma$-field $\cH\supseteq\cF$ write $c^{k}_{\cH}=\E[Y^{k}\mid\cH]-\E[Y^{k}\mid\cF]$. Besides $\cH=\cG$, a reader of the full text, we use $\cH=\cF\vee\sigma(x)$, a reader who sees only the embedding.

\begin{proposition}[Anchored concepts]\label{prop:concept}
Let $Y=(Y^{1},\dots,Y^{K})$ be square integrable and $\cH\supseteq\cF$.
\begin{enumerate}[label=(\roman*)]
\item Uniqueness. $c^{k}_{\cH}$ is the almost surely unique minimizer of $\E\big[(Y^{k}-\E[Y^{k}\mid\cF]-c)^{2}\big]$ over $c\in L^{2}(\cH)$.
\item Units and adding up. For $a\in\mathbb{R}^{K}$, the concept of the outcome $a^{\top}Y$ is $a^{\top}c_{\cH}$. If $a^{\top}Y$ is $\cF$-measurable, then $a^{\top}c_{\cH}=0$.
\item Invariance. If $\cH=\cF\vee\sigma(x)$ and $\chi:\mathbb{R}^{J}\to\mathbb{R}^{J}$ is a bijection that is measurable with measurable inverse, then replacing $x$ by $\chi(x)$ leaves every $c^{k}_{\cH}$ unchanged.
\end{enumerate}
\end{proposition}

Because the concept solves a least-squares problem in observables, it can be estimated, and because the solution is unique, there is nothing to rotate (part (i)). By part (iii), a change of coordinates changes the embedding and leaves every concept unchanged. Part (ii) is the reason the choice of outcomes matters: a linear identity among outcomes holds among their concepts with the same coefficients. A text that tells the reader nothing new has a concept of zero for every outcome, whatever it says: if $w$ is $\cF$-measurable, then $\cG=\cF$.

\subsection{An Exact Return Identity}\label{sec:fates}

\begin{figure}[htbp]
\centering\includegraphics[width=\textwidth]{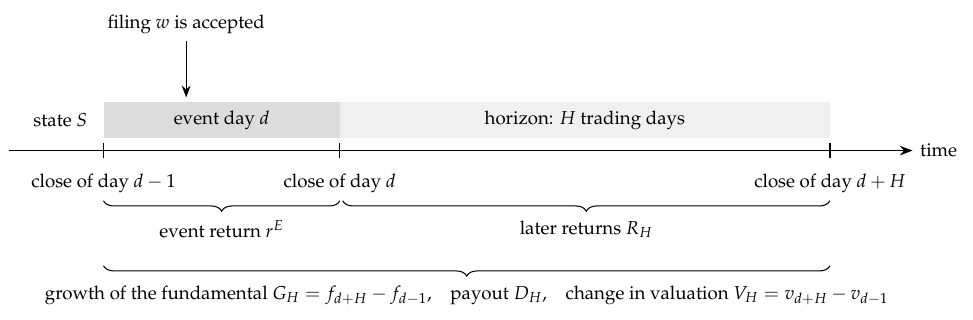}
\caption{Timing}\label{fig:timeline}
\floatnote{Information before the window is summarized by the state $S$ at the close of day $d-1$. The event window is the first trading day $d$ whose close follows the acceptance of the filing. Later returns, the growth of the fundamental, the payout, and the change in valuation are measured to the close $H$ trading days after the window.}
\end{figure}

Let $P_{s}>0$ be the firm's stock price and $\Div_{s}\ge0$ the dividend paid in period $s$. Write $p_{s}=\log P_{s}$, $r_{s}=\log\big((P_{s}+\Div_{s})/P_{s-1}\big)$ for the log return and $\delta_{s}=\log(1+\Div_{s}/P_{s})$ for the log dividend yield. Let $f_{s}$ be any real-valued process, the fundamental, and $v_{s}=p_{s}-f_{s}$ the valuation ratio. In the application $f$ is the log of book equity or of sales per share. The window runs from the close of period $\tau_{0}$ to the close of period $\tau_{1}$. For a horizon of $H\ge1$ periods after the window, define
\begin{align*}
r^{E}&=\sum_{s=\tau_{0}+1}^{\tau_{1}}r_{s}, & R_{H}&=\sum_{s=\tau_{1}+1}^{\tau_{1}+H}r_{s}, & G_{H}&=f_{\tau_{1}+H}-f_{\tau_{0}},\\
D_{H}&=\sum_{s=\tau_{0}+1}^{\tau_{1}+H}\delta_{s}, & V_{H}&=v_{\tau_{1}+H}-v_{\tau_{0}}. &&
\end{align*}
These are the event return, the later returns, the growth of the fundamental, the payout, and the change in valuation (Figure~\ref{fig:timeline}).

For an outcome $Y$ write $\tilde Y=Y-\E[Y\mid\cF]$ for its unexpected part. The unexpected event return splits into the part that the text accounts for and a remainder,
\[
\tilde r^{E}=c^{r}+u^{r},\qquad c^{r}=\E[r^{E}\mid\cG]-\E[r^{E}\mid\cF],\qquad u^{r}=r^{E}-\E[r^{E}\mid\cG],
\]
and the two are uncorrelated because $\E[u^{r}\mid\cG]=0$. We call $c^{r}$ the text part of the event return, $u^{r}$ the remainder, and $\tilde r^{E}$ the whole move. More generally, the text part of an outcome is its anchored concept, and its text share is the fraction of its unexpected variance that the text accounts for, $\Var(c^{k})/\Var(\tilde Y^{k})$. For each of these components $a\in\{c^{r},u^{r},\tilde r^{E}\}$ with positive variance, define at horizon $H$
\begin{align*}
\omega_{G}(a)&=\frac{\Cov(a,\tilde G_{H})}{\Var(a)}, & \omega_{D}(a)&=\frac{\Cov(a,\tilde D_{H})}{\Var(a)},\\
\omega_{R}(a)&=-\frac{\Cov(a,\tilde R_{H})}{\Var(a)}, & \omega_{V}(a)&=\frac{\Cov(a,\tilde V_{H})}{\Var(a)}.
\end{align*}
We refer to these as the growth, payout, offset, and valuation shares, and together as the shares of the identity. The offset share is positive when later returns reverse the move and negative when they extend it.

\begin{proposition}[Return identity]\label{prop:fates}
\begin{enumerate}[label=(\alph*)]
\item For every $H\ge1$ and along every path, $r^{E}=G_{H}+D_{H}-R_{H}+V_{H}$.
\item Suppose the five terms are square integrable. Then $c^{r}=c^{G}_{H}+c^{D}_{H}-c^{R}_{H}+c^{V}_{H}$, and for each component $a\in\{c^{r},u^{r},\tilde r^{E}\}$ with $\Var(a)>0$,
\[
\omega_{G}(a)+\omega_{D}(a)+\omega_{R}(a)+\omega_{V}(a)=1\qquad\text{at every horizon }H.
\]
For $a=c^{r}$ the shares can be written with concepts alone: $\omega_{G}(c^{r})=\Cov(c^{r},c^{G}_{H})/\Var(c^{r})$, and likewise for the other three.
\item Let $\bar r_{s}$ be the log return on a benchmark portfolio in period $s$, and let $\bar r^{E}$ and $\bar R_{H}$, both square integrable, be its sums over the window and over the $H$ periods after it. Parts (a) and (b) hold with $r^{E}$, $R_{H}$, and $V_{H}$ replaced by $r^{E}-\bar r^{E}$, $R_{H}-\bar R_{H}$, and $V_{H}-\bar r^{E}-\bar R_{H}$.
\end{enumerate}
\end{proposition}

By horizon $H$, therefore, part of a price move has been realized in fundamentals, as growth or as payout; part has been offset by later returns or, when the offset share is negative, extended by them; and the rest is still in the valuation ratio. The identity is exact: unlike the log-linear identity of \citet{campbell1988dividend}, it requires no approximation and no dividends, which matters for daily data and for the many firms that pay none \citep{bendavid2026crimes}. Every term is observable, so no term is obtained as a residual, as one is in decompositions based on vector autoregressions. The identity holds at every finite horizon; the change in valuation absorbs whatever has been neither realized nor offset by $H$. Part (c) covers benchmark-adjusted returns, which we use in the empirical work: the change in valuation is then measured net of the benchmark's cumulative return.

If the valuation share vanishes as the horizon grows, a move is eventually divided between the growth share and the sum of the offset and payout shares. Returns net of payouts are price changes, so this sum is the part of the move that later price changes offset. The two parts are the counterparts of cash-flow news and discount-rate news in the decomposition of \citet{campbell1991variance}, with an observed fundamental in place of dividends. In the example below the correspondence is exact for discount-rate news and for news about current cash flows, and approximate for news about later growth, which also lowers later payout yields (Proposition~\ref{prop:exshares}). At a finite horizon the shares need care. A persistent fall in the discount rate raises the price today and lowers returns a little in every later period, so after a month or a quarter little of it has been offset, and a small offset share does not show that a text carries no discount-rate news. A reaction that is too small or too large is corrected by later returns, which enter the offset share with their own sign. The example that follows makes both points explicit.

\subsection{An Example: Three Kinds of News}\label{sec:example}

Time is discrete, and $\cF_{s}$ denotes public information at the end of period $s$. An event occupies one period, from $t$ to $t+1$, so that $\cF=\cF_{t}$ and $\Fp=\cF_{t+1}$. Let $d_{s}$ be the log dividend and $\pd_{s}$ the log price-dividend ratio, and suppose that returns obey $r_{s+1}=k_{0}+\rho\,\pd_{s+1}-\pd_{s}+\Delta d_{s+1}$ with $0<\rho<1$, the log-linear relation of \citet{campbell1988dividend}, treated as exact for the purpose of the example. Dividend growth and required returns have persistent components,
\[
\Delta d_{s+1}=g_{s}+\varepsilon_{s+1},\qquad g_{s+1}=\phi_{g}g_{s}+\eta_{s+1},\qquad \lambda_{s+1}=\phi_{\lambda}\lambda_{s}+\xi_{s+1},
\]
with $|\phi_{g}|,|\phi_{\lambda}|<1$. The market values the firm as if expected growth were $\hat g_{s}=g_{s}+\beta_{s}$ and discounts at $\lambda_{s}$:
\begin{equation}\label{eq:pd}
\pd_{s}=k_{1}+a_{g}\hat g_{s}-a_{\lambda}\lambda_{s},\qquad a_{g}=\frac{1}{1-\rho\phi_{g}},\quad a_{\lambda}=\frac{1}{1-\rho\phi_{\lambda}}.
\end{equation}
This is the log-linear present value when, up to constants, the market expects dividend growth $\phi_{g}^{j}\hat g_{s}$ and returns $\phi_{\lambda}^{j}\lambda_{s}$ in period $s+1+j$, for $j\ge0$. The wedge $\beta$ is the market's error about growth. The text published in period $s+1$ conveys part of that period's shocks. Write $\cG_{s+1}=\cF_{s}\vee\sigma(w_{s+1})\subseteq\cF_{s+1}$ and let $(\varepsilon^{w}_{s+1},\eta^{w}_{s+1},\xi^{w}_{s+1})$ be the expectations of $(\varepsilon_{s+1},\eta_{s+1},\xi_{s+1})$ given $\cG_{s+1}$. The market's reaction to the growth news in the text is governed by a coefficient $\theta>0$:
\begin{equation}\label{eq:beta}
\beta_{s+1}=\phi_{\beta}\beta_{s}-(1-\theta)\,\eta^{w}_{s+1}+\varsigma_{s+1},\qquad |\phi_{\beta}|<1.
\end{equation}
The reaction is complete when $\theta=1$, too small when $\theta<1$, and too large when $\theta>1$; errors are corrected at rate $1-\phi_{\beta}$. We assume that the states $(g_{s},\lambda_{s},\beta_{s})$ are $\cF_{s}$-measurable, that $(\varepsilon_{s},\eta_{s},\xi_{s},\varsigma_{s})$ are martingale differences with respect to $\{\cF_{s}\}$ with finite second moments, and that $\E[\varsigma_{s+1}\mid\cG_{s+1}]=0$.

Take $f=d$, so that the valuation ratio is $\pd$. Then $G_{H}=d_{t+1+H}-d_{t}$, $R_{H}=\sum_{j=1}^{H}r_{t+1+j}$, and $V_{H}=\pd_{t+1+H}-\pd_{t}$, and the payout term of the identity, $D_{H}=r_{t+1}+R_{H}-G_{H}-V_{H}$, is the sum of the log-linear dividend yields $k_{0}-(1-\rho)\pd_{s}$ over $s=t+1,\dots,t+1+H$. For the event at $t+1$ we drop the time subscript and write $(\varepsilon^{w},\eta^{w},\xi^{w})$.

\begin{proposition}[Concepts in the example]\label{prop:example}
In the example, for every $H\ge1$:
\begin{enumerate}[label=(\alph*)]
\item $c^{r}=\varepsilon^{w}+\rho\,\theta\,a_{g}\,\eta^{w}-\rho\,a_{\lambda}\,\xi^{w}$;
\item the concepts of the later terms are
\begin{align*}
c^{G}_{H}&=\varepsilon^{w}+\frac{1-\phi_{g}^{H}}{1-\phi_{g}}\,\eta^{w},\\
c^{R}_{H}&=\frac{1-\phi_{\lambda}^{H}}{1-\phi_{\lambda}}\,\xi^{w}+(1-\theta)(1-\rho\phi_{\beta})\,a_{g}\,\frac{1-\phi_{\beta}^{H}}{1-\phi_{\beta}}\,\eta^{w},\\
c^{V}_{H}&=a_{g}\big(\phi_{g}^{H}-(1-\theta)\phi_{\beta}^{H}\big)\eta^{w}-a_{\lambda}\phi_{\lambda}^{H}\,\xi^{w}.
\end{align*}
\end{enumerate}
\end{proposition}

Each kind of news leaves its own signature. News about current cash flows, $\varepsilon^{w}$, is realized at once: it enters the growth of the fundamental in full at every horizon and never enters valuation. News about persistent growth, $\eta^{w}$, enters valuation and is realized in the fundamental at the rate at which the growth it announces arrives. When the reaction is complete, its coefficient in the change in valuation is $a_{g}\phi_{g}^{H}$, which for $0<\phi_{g}<1$ decays geometrically with a half-life of $\log 2/|\log\phi_{g}|$ periods. Discount-rate news, $\xi^{w}$, is never realized in the growth of the fundamental; it leaves valuation only as later returns offset it. A reaction that is too small or too large appears in later returns with a definite sign. Suppose that the three kinds of news are uncorrelated and that the text carries growth news and no discount-rate news. Then the offset share of the text part has the sign of $\theta-1$ at every horizon, and it is zero at every horizon if and only if $\theta=1$. The absence of drift after news corresponds to an offset share of zero, and a negative offset share measures the drift per unit of the initial move. Proposition~\ref{prop:exshares} gives the four shares in closed form.

\subsection{Estimation and Scope}\label{sec:est}

All of the quantities above are variances and covariances of conditional expectations. We estimate each as the sample mean of an orthogonal score evaluated at cross-fitted expectations \citep{chernozhukov2018double,williamson2021nonparametric}; Proposition~\ref{prop:score} states the result. The error from using fitted expectations is then the product of two estimation errors. For a cross moment one accurate fit is therefore enough, and the event return, which enters every share, is the best-measured outcome. For a concept's own variance the two errors coincide: the estimate falls short of the truth by the mean squared error of the fitted concept, so a text share is understated. A share of the text part has this variance in its denominator. Its distance from the corresponding share of the remainder is overstated by a factor of about $\Var(c)/(\Var(c)-\mathrm{MSE})$, where $c$ is the text part and MSE is the mean squared error of its fit, and, under the approximation stated in Appendix~\ref{app:proofs}, a test that the two shares are equal remains correctly centered. Section~\ref{sec:estimator} describes how the folds and the standard errors allow for dependence across firms and over time, and Appendix~\ref{app:mc} checks the procedure by simulation.

A readout is fitted within a class, and the class may not contain the true conditional expectation. The estimator then recovers the variances and covariances of the projection of each concept on the class, and when the same class is used for every outcome the adding-up identities hold for these projections (Proposition~\ref{prop:readable}). The shares describe the component of the text that the class captures, and two classes may capture components with different shares. In the example, a class that captures only the news about current cash flows can yield a text part whose growth share is one, although the text also carries news that stays in valuation. A projected concept is, moreover, unchanged by a change of coordinates only if the change leaves the class unchanged. We therefore state the class used for every result, choose it by out-of-sample fit, and report in Section~\ref{sec:fateresults} how the shares change with a more flexible class.

\section{Data and Measurement}\label{sec:data}

\subsection{Filings}

The texts are current reports on Form 8-K filed with the Securities and Exchange Commission. A firm must file a current report within four business days of a material event, such as the release of quarterly results, the signing of a material agreement, or the departure of an officer, and the Commission's item numbers classify the event \citep{lerman2010new}. We download from the Commission's Electronic Data Gathering, Analysis, and Retrieval (EDGAR) system every Form 8-K and 8-K/A filed between January 2014 and December 2023 by a firm with a common stock in the database of the Center for Research in Security Prices (CRSP), matched to Compustat by the CRSP--Compustat link on the filing date; the sample starts in 2014 because the training text of our encoder ends in December 2013. There are 465,826 such filings. We obtain the text of all but 24, and each carries the time, to the second, at which EDGAR accepted it.

The text of a filing is the body of the report, without its cover page and signature block, followed by any press release furnished as Exhibit 99. A press release accompanies 63\% of the filings and 98\% of those that report quarterly results. The unit of observation is a firm-day, and the main sample consists of the firm-days to which at least one filing is assigned by the rule described below. We require a stock price of at least \$1 at the close before the event day, which leaves 416,532 firm-days; 4\% of them have more than one filing. We use the first three months of 2014 only to build each firm's filing history, so the main sample has \N{nA} firm-days for \N{nFirms} firms.

We classify a firm-day by the items of its filings. The most common are results of operations (Item 2.02, 32\% of firm-days), changes of officers and directors (17\%), other events (Item 8.01, 14\%), Regulation FD disclosures (12\%), governance matters and shareholder votes (8\%), material agreements (7\%), and financing (7\%). Unlike the labels that a language model or a dictionary attaches to a text, these categories are assigned by the filer under the Commission's rules and are known when the filing is read. An earnings announcement recorded by the Institutional Brokers' Estimate System (I/B/E/S) falls on 28\% of the event days defined below. We call these announcement days and the rest other event days.

\subsection{The Event Window}

The acceptance time fixes the event window. A filing accepted at or before 4 p.m.\ Eastern time on a trading day is assigned to that day, and any other filing to the next trading day. The event window is this day $d$, from the close of day $d-1$ to the close of day $d$, and all horizons are counted from the close of day $d$. We call day $d$ the event day and its unexpected return the event-day move. Filings are accepted after the close on 59\% of firm-days, before the open on 26\%, and during trading hours on 13\%; on the remaining 2\% they are accepted in more than one of these periods or on a day without trading. For releases of quarterly results we use the announcement time recorded by I/B/E/S when it places the release one trading day before the filing does, which is the case for 4\% of them. Horizons are $H=1$, 5, 21, 63, 252, and 504 trading days, which we call a day, a week, a month, a quarter, a year, and two years, with 756 days (three years) added in Table~\ref{tab:long}. Returns are available through December 2025.

The acceptance time is the latest moment at which the text could have become public. A firm may issue a press release some hours before it furnishes the release on Form 8-K, and the price can then respond before the filing is accepted. A response earlier on the same trading day is inside the window; Section~\ref{sec:share} reports the text share separately for the two parts of the event day. Assumption~\ref{ass:pit} requires only that the text be public by the end of the window, which the acceptance time guarantees.

\subsection{Language Model}

The encoder is the PIT-4B model of \citet{kelly2026scaling}, a point-in-time language model with about four billion parameters, in the publicly released checkpoint whose training text ends in December 2013 (PIT-4B-201312). We computed the embeddings in November 2026. The same weights are used for every year of the sample, so that Assumption~\ref{ass:pit} holds and all embeddings share one coordinate system. A filing is represented by the average of the model's final-layer representations of its first 2,048 tokens, a vector of dimension 4,096, and a firm-day by the average over its filings. The median filing has 1,718 tokens, and 45\% of filings reach the limit.

Much of a current report repeats the language of the firm's earlier reports, and text that tells the reader nothing new has a concept of zero (Section~\ref{sec:concepts}). A linear readout cannot discard such content by itself. We therefore subtract from each firm-day vector an exponentially weighted average of the same firm's vectors on earlier event days, and we include the leading principal components of that average in the state. We call the result the novelty embedding and use it as $x$ throughout.

\subsection{Outcomes and State}\label{sec:outcomes}

Returns, prices, dividends, and volume are from CRSP. The event return is the log return from the close of day $d-1$ to the close of day $d$. Market-adjusted returns are net of the log return on the value-weighted market, and the change in valuation is adjusted accordingly, as in Proposition~\ref{prop:fates}(c). Returns include delisting returns, and a delisted position is held in cash thereafter.

Fundamentals are from the Compustat quarterly file. Sales are summed over the trailing four quarters, and book equity is stockholders' equity plus deferred taxes minus preferred stock. For the accounting of Proposition~\ref{prop:fates}, what matters is the moment at which a quarter's results become known. We treat a quarter as known from the close of the first trading day whose close follows the release of its results. The time of the release is taken from I/B/E/S when its date matches the report date in Compustat; otherwise the quarter is dated on the trading day after the report date. Results released after the close of day $d-1$ are thus part of the news of day $d$ and not of what was known before it. The fundamental before the window is the most recent positive value known at the close of day $d-1$, and the fundamental at horizon $H$ is the most recent positive value known at the close of day $d+H$. The valuation ratio is the log of market capitalization over the fundamental, which equals the log of the price over the fundamental per share; $f$ is thus the log fundamental per share and $G_{H}$ its growth. The sales sample has \N{nVS} firm-days and the book-equity sample \N{nBook}.

Two features of the timing matter for what follows. Because a quarter enters the fundamental when its results become known, the growth of the fundamental after an announcement day includes the quarter just announced, at every horizon. And because sales are a sum over four quarters, the growth measured after a year mixes quarters of different vintages. The growth of trailing sales between two dates is the sum of the changes that occur as each new quarter becomes known. We group the quarters that become known between the close of day $d-1$ and the close of day $d+H$ by when they were earned: the quarter announced on the event day; quarters that had ended before the event day and were reported after it; the quarter in progress on the event day; and quarters that began after it. With the change in the number of shares, the five pieces sum exactly to the growth of sales per share.

The state $S$ contains what is known at the close of day $d-1$: firm size and stock price, past returns at five horizons, volatility, turnover, illiquidity, dividend yield, institutional ownership, recent filing and analyst activity, the projected timing of the next earnings announcement, industry, market returns and volatility, and the leading principal components of the firm's past text. For the outcomes that involve fundamentals it also contains valuation, profitability, investment, and leverage ratios, the level and dispersion of analysts' earnings forecasts, analyst coverage, the return implied by outstanding price targets, and option-implied volatility. Analyst data and the dates and times of earnings announcements are from I/B/E/S, and implied volatility is from OptionMetrics.

We date every variable in the state at or before the close of day $d-1$, with two qualifications. The accounting values, which enter the state only for the outcomes that involve fundamentals, come from the standard Compustat file and may have been restated after their release. And the directions that summarize a firm's past text are estimated on the filings of 2014, which affects only the event days of that year and none of the chronological results. The state does not use the realized date of any earnings announcement. It uses a projected date, 52 weeks after an announcement of the previous year, which falls within five trading days of the actual date for 83\% of announcements. Accounting ratios use only quarters known under the rule above, analysts' forecasts are those time-stamped before 4 p.m.\ on day $d-1$, and a variable's rank in the cross section is taken against the firm-days of the three preceding months. Appendix~\ref{app:est} lists the variables, their timing, and the functional form. Appendix~\ref{app:vars} defines the outcomes and reports summary statistics (Tables~\ref{tab:defs} and~\ref{tab:sumstats}).

As a benchmark we use days without news: \N{nNoNews} firm-days drawn at random from those on which the firm had filed a current report during the preceding quarter but, on days $d-1$, $d$, and $d+1$, neither filed one nor appeared in the news collection of \citet{dong2024fnspid}.

\subsection{Readout, Estimator, and Inference}\label{sec:estimator}

Estimation follows Section~\ref{sec:est}. The state is partialed out of the outcomes and the embedding, and the readout is a ridge regression of the residual outcome on the residual embedding, linear in the novelty embedding. A pooled linear readout would give every firm the same response to a given text. We let the scale of the response differ across nine groups formed by size third and by the kind of event day: a day with a filing that reports quarterly results, another day with a press release, and a day without one. Each group has one multiplicative factor, the slope of the unexpected outcome on the fitted concept within the group, and we call the nine factors group scales. A second possible adjustment makes the response proportional to the firm's volatility over the previous 63 days. We call a choice of adjustments a design and choose it on early data, as described below.

For two outcomes $Y$ and $Z$, let $\tilde Y$ and $\tilde Z$ be their unexpected parts given the state and $\hat c^{Y}$ and $\hat c^{Z}$ the fitted concepts. The covariance of the two concepts is estimated by the sample mean of the score
\[
\psi_{YZ}=\tilde Y\,\hat c^{Z}+\tilde Z\,\hat c^{Y}-\hat c^{Y}\hat c^{Z},
\]
and the variance of a concept by the mean of $\psi_{YY}=2\tilde Y\hat c^{Y}-(\hat c^{Y})^{2}$. Fitted values for a firm-day come from readouts estimated without it, so an error in the readouts enters the mean of $\psi_{YZ}$ only through the product of the errors in $\hat c^{Y}$ and $\hat c^{Z}$ (Proposition~\ref{prop:score}). Every share, contribution, and test statistic below is a ratio, or a difference of ratios, of such means and of the second moments of the unexpected outcomes, and its standard error follows from the joint covariance of these means by the delta method. Appendix~\ref{app:est} gives the formulas.

We fit the readouts in two ways, which answer different questions. The baseline fit is cross-fitted: fitted values for a calendar quarter come from estimates that exclude that quarter and, for an outcome measured over $H$ days, the months on either side whose outcome windows could overlap it. We call the excluded months the gap. The same gap is needed for the readout of the event return whenever it is combined with an outcome at horizon $H$. A firm's later returns include the event-day returns of its next filings, whose text resembles that of the current one, so a readout of the event return fitted on the months that follow is fitted, in part, to the outcome with which it is compared. For each horizon we therefore fit a readout of the event return that excludes the same months as the outcomes at that horizon, in the manner of the purged samples of \citet{lopezdeprado2018advances}. The gap grows with the horizon, from 2 months for a week to 26 months for two years, and the readout loses precision accordingly: for all firm-days the text share of the event return is \N{tsB}\% without a gap, \N{kTsw}\% with the gap of the one-week outcome, and \N{tsBgapY}\% with that of the one-year outcome (the last figure is untabulated). The baseline fit uses months after the quarter as well as before it. It describes how text and returns were related over the decade.

The chronological fit uses, for each quarter, only what was available before it. The readout of each later outcome is fitted on earlier months, less the gap, and the readout of the event return on earlier months with no gap. The ridge penalty is the one with the best out-of-sample record over the earlier quarters whose outcomes had been realized, and the group scales come from earlier months. The design is chosen in the same spirit: of the four combinations of group scales and the volatility adjustment, we use in both fits the one with the best chronological fit of the event return in 2015--2016, which has group scales and no volatility adjustment (Table~\ref{tab:design}). We report chronological results for 2017--2023. They describe what a researcher could have computed at the time, and Appendix~\ref{app:est} lists the data that each step of this fit uses.

In both fits the penalty, the group scales, and the excluded months are specific to each outcome, so the estimated contributions of the text part are not constrained to add up to its text share. With sales as the fundamental, the four contributions sum to the text share to within \N{addB} percentage points in the baseline fit and \N{addC} in the chronological fit (Table~\ref{tab:addup}). The shares of the whole move sum to one to within 0.0001.

Our panel has repeated firms, shocks common to many firms, and outcomes whose windows overlap, whereas Proposition~\ref{prop:score} is stated for independent observations. The inference allows for these features in four ways, which Appendix~\ref{app:est} describes in detail. The first is the gap. The second is the standard errors, which combine clustering by firm with the estimator of \citet{driscoll1998consistent} for dependence across firms on the same day and across nearby days, with a lag length equal to the horizon of the outcome, capped at 300 trading days, in the manner of \citet{thompson2011simple}. The third is a set of checks that do not rest on the same asymptotic argument: a bootstrap over blocks of calendar quarters, a jackknife that deletes one calendar quarter at a time, and a placebo in which embeddings are reassigned at random across firms within each day, where the estimated text share of the event return is $\N{tsPerm}\%$ (standard error \N{tsPermSE}). The fourth is simulation. In panels simulated from the example of Section~\ref{sec:example}, with common shocks, fat tails, stale content, and uninformative coordinates, the mean estimates of the text share and of the offset shares are within 5\% of the true values for the component that a linear readout captures, except in the setting with the least informative text, in which the text accounts for 0.4\% of return variance and the largest bias is 8\%; nominal 95\% confidence intervals cover the true values in 90.0\% to 96.5\% of replications (Appendix~\ref{app:mc}). A decade holds about ten nonoverlapping one-year periods and five two-year periods, so the standard errors at these horizons rest on fewer independent observations than those at shorter horizons; Section~\ref{sec:robust} reports the checks that bear on them.

\section{The Text Share of Event Returns}\label{sec:share}

Table~\ref{tab:share} reports the text share of the event return. In the baseline fit the text of the filings accounts for \N{tsB}\% of the variance of the market-adjusted return on the event day. In the chronological fit for 2017--2023 it accounts for \N{tsC}\%; the baseline figure for the same years is \N{tsBlate}\% (Section~\ref{sec:robust}). Identifiable news has long been found to explain a modest part of individual stock returns \citep{roll1988r2}; the figure here is an out-of-sample share for a single day's filings.

\begin{table}[tbp]
\caption{Text Share of Event-Return Variance}\label{tab:share}
\tabwrap{\begin{tabular}{@{}lrrrr@{}}
\toprule
 & \multicolumn{2}{c}{Baseline, 2014--2023} & \multicolumn{2}{c}{Chronological, 2017--2023} \\
\cmidrule(lr){2-3}\cmidrule(l){4-5}
Sample and Return & Text Share (\%) & SE & Text Share (\%) & SE \\
\midrule
\multicolumn{5}{@{}l}{\textit{Panel A. All firm-days}} \\
Market-adjusted return & 5.49 & 0.18 & 4.71 & 0.17 \\
Raw return & 5.29 & 0.18 & 4.51 & 0.18 \\
\quad Previous close to open & 6.81 & 0.22 & 5.89 & 0.21 \\
\quad Open to close & 0.43 & 0.04 & 0.37 & 0.05 \\
\addlinespace[3pt]\multicolumn{5}{@{}l}{\textit{Panel B. Market-adjusted return, by firm size}} \\
Smallest third & 5.83 & 0.27 & 5.08 & 0.26 \\
Middle third & 5.75 & 0.27 & 4.95 & 0.25 \\
Largest third & 3.30 & 0.18 & 2.57 & 0.16 \\
\addlinespace[3pt]\multicolumn{5}{@{}l}{\textit{Panel C. Market-adjusted return, by content of the filing}} \\
With a press release & 6.18 & 0.20 & 5.33 & 0.20 \\
Without a press release & 1.55 & 0.23 & 1.50 & 0.19 \\
Announcement days & 5.73 & 0.23 & 5.00 & 0.25 \\
Other event days & 5.33 & 0.25 & 4.54 & 0.22 \\
\addlinespace[3pt]\multicolumn{5}{@{}l}{\textit{Panel D. Market-adjusted return, by time of acceptance}} \\
Before the open & 8.15 & 0.29 & 6.60 & 0.27 \\
During trading hours & 2.43 & 0.56 & 2.63 & 0.49 \\
After the close & 3.69 & 0.23 & 3.36 & 0.22 \\
\addlinespace[3pt]\multicolumn{5}{@{}l}{\textit{Panel E. Raw return, by part of the day and time of acceptance}} \\
Before the open: previous close to open & 10.16 & 0.37 & 8.25 & 0.34 \\
Before the open: open to close & 0.64 & 0.08 & 0.53 & 0.08 \\
During trading hours: previous close to open & 0.53 & 0.98 & 1.53 & 0.80 \\
During trading hours: open to close & 0.42 & 0.13 & 0.39 & 0.11 \\
After the close: previous close to open & 4.02 & 0.28 & 3.71 & 0.26 \\
After the close: open to close & 0.32 & 0.05 & 0.28 & 0.05 \\
\bottomrule
\end{tabular}}
\floatnote{The table reports the text share of the event return: the variance of the text part as a percentage of the variance of the unexpected return on the event day. The baseline fit is cross-fitted over calendar quarters of 2014--2023 (\N{nA} firm-days); the chronological fit uses earlier months only and is reported for 2017--2023 (292,427 firm-days). Size thirds are thirds of market capitalization relative to the firm-days of the three preceding months. Announcement days are event days on which the Institutional Brokers' Estimate System (I/B/E/S) records an earnings announcement. The last three rows of Panel A and all of Panel E use raw returns; the other rows use market-adjusted returns. SE is the standard error, which allows for firm clusters and for common shocks across firms on the same day.}
\end{table}

The share varies across firm-days. It is larger for small firms (\N{tsBsmall}\% in the smallest third against \N{tsBbig}\% in the largest) and for filings that furnish a press release (\N{tsBrel}\% against \N{tsBnorel}\%). Announcement days and other event days differ little, at \N{tsBea}\% and \N{tsBnoea}\%. By the Commission's items, the share is highest for material agreements (\N{tsTyAgree}\%), followed by results of operations (\N{tsTyEarn}\%) and other events (\N{tsTyOther}\%); for changes of officers, governance matters, and acquisitions the fitted readout explains none of the variance (Table~\ref{tab:types}).

\subsection{When the Price Responds}

The share is concentrated in the return from the previous close to the open. The text accounts for \N{onB}\% of the variance of that overnight return and for \N{idB}\% of the variance of the return from the open to the close. For filings accepted before the open the two figures are \N{onPre}\% and \N{idPre}\%, and for filings accepted after the close they are \N{onPost}\% and \N{idPost}\%. When a filing arrives outside trading hours, the text part is largely in the price by the next open, in keeping with the finding of \citet{wu2025extracting} for earnings press releases. For filings accepted during trading hours, 13\% of firm-days, the event day mixes returns from before and after the filing. The text share of the day's market-adjusted return is \N{tsBreg}\% for these filings. The share of the overnight return, which precedes them, is \N{onReg}\% (standard error \N{onRegSE}), and that of the return from the open to the close is \N{idReg}\% (standard error \N{idRegSE}).

The acceptance time matters for the choice of day. If the window is keyed to the calendar date of acceptance, as it must be when only dates are known, the response to filings accepted after the close falls on the following day: the text then accounts for \N{dtNext}\% of the variance of the next day's return (Table~\ref{tab:texts}). With such a window the day-after return would look predictable from the text.

\subsection{How Much Text}

A reader of the title of a filing, or of its opening, knows less than a reader of the whole, and Proposition~\ref{prop:info} measures the loss in the units of the outcome. The title accounts for \N{tsTitle}\% of the variance of the event return, the first 256 tokens for \N{tsLead}\%, and the first 2,048 tokens for \N{tsB}\% (Table~\ref{tab:texts}). Two-thirds of what the text accounts for is lost by reading only the title. Each text has its own embedding and readout, so the comparison is between projected concepts in the sense of Proposition~\ref{prop:readable}; the ordering agrees with the nesting of the three texts. Using the novelty embedding in place of the unadjusted one raises the share from \N{tsRaw}\% to \N{tsB}\%: language that a firm repeats from one report to the next has a concept of zero, and a linear readout does better once it is taken out. A more flexible class changes little (Section~\ref{sec:robust}).

\section{Decomposing Event-Day Price Moves}\label{sec:fateresults}

\begin{figure}[tbp]
\centering\includegraphics[width=\textwidth]{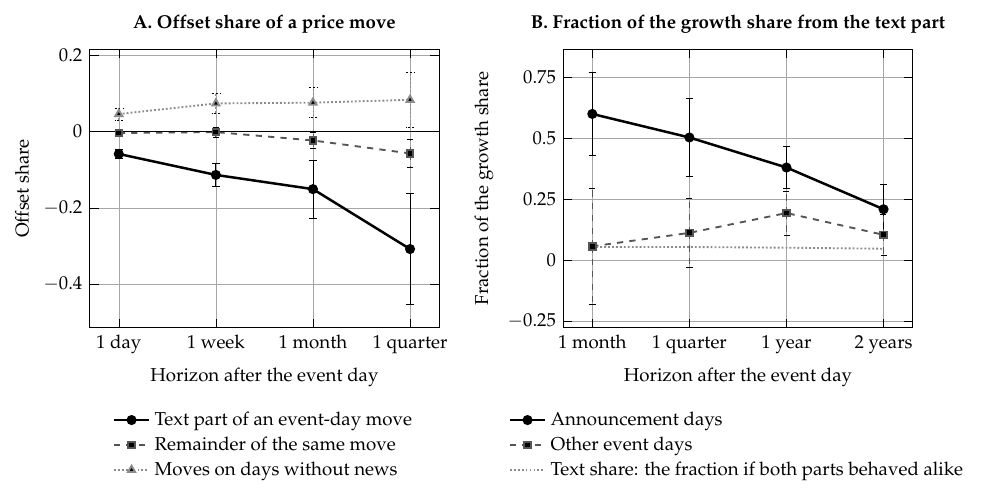}
\caption{What Follows an Event-Day Price Move, by Horizon}\label{fig:fates}
\floatnote{The figure shows the offset share of three kinds of price move (Panel A) and the fraction of the growth share that the text part contributes (Panel B). The offset share is the regression slope of the market-adjusted return over the horizon on the move, with the sign reversed; a negative value means that later returns extend the move. The levels for the text part are displaced away from those of the remainder (Section~\ref{sec:est}). Days without news are firm-days on which the firm neither filed a current report nor appeared in the news collection of \citet{dong2024fnspid}. In Panel B the fundamental is sales per share, and the dotted line is the text share for all event days, the fraction the text part would contribute if it and the remainder behaved alike. Estimates are from the baseline fit, 2014--2023. Bars are 95\% confidence intervals.}
\end{figure}

Table~\ref{tab:contrib} reports the accounting of Proposition~\ref{prop:fates} for the text part of an event-day move, $c^{r}$, and for the whole move. Figure~\ref{fig:fates} and Table~\ref{tab:levels} add the remainder of the event-day move, $u^{r}$, and the unexpected return on a day without news. Returns are market adjusted, and the fundamental is sales per share. In the baseline fit, every quantity that involves the text part uses the readout of the event return fitted with the gap of its horizon (Section~\ref{sec:estimator}).

The text part and the remainder are uncorrelated, so the share of the whole move is the variance-weighted average of their shares. For any later outcome $Y$,
\[
\omega(\tilde r^{E})=\frac{\Cov(c^{r},\tilde Y)}{\Var(\tilde r^{E})}+\frac{\Cov(u^{r},\tilde Y)}{\Var(\tilde r^{E})},
\]
and the first term equals the text share of the event return times the share of the text part. We call it the contribution of the text part to the share of the whole move. Table~\ref{tab:contrib} reports the contributions of the text part in percent of the variance of the whole move, together with the fraction of the whole move's share that the text part carries. If the text part and the remainder had equal shares, that fraction would equal the text share of the event return, about one-twentieth. Unless we say otherwise, a $z$-statistic below is the difference between a contribution and the proportional one, the text share times the share of the whole move, divided by its standard error.

Contributions are less exposed than shares to the error of the fitted readouts. Their denominator is the variance of the whole move, which involves no readout, and the error enters only through the numerator: under the approximation of Appendix~\ref{app:proofs}, it subtracts the readout's mean squared error times the remainder's share, a negligible amount for the one-week offset, where that share is near zero, and a small understatement of the sales contributions. The shares of the text part are subject to the displacement of Section~\ref{sec:est}, most of all at long horizons, where the gap makes the fit less accurate.

\begin{table}[tbp]
\caption{Contribution of the Text Part to the Shares of the Whole Event-Day Move}\label{tab:contrib}
\tabwrap{\setlength{\tabcolsep}{3.4pt}\begin{tabular}{@{}lrrrrrrrrrrr@{}}
\toprule
 & \multicolumn{6}{c}{Baseline, 2014--2023} & \multicolumn{5}{c}{Chronological, 2017--2023} \\
\cmidrule(lr){2-7}\cmidrule(l){8-12}
 & Text & Whole & Text & & & & Whole & Text & & & \\
Horizon & Share (\%) & Move & Part & SE & Fraction & $z$ & Move & Part & SE & Fraction & $z$ \\
\midrule
\multicolumn{12}{@{}l}{\textit{Panel A. Later returns (a negative entry means that later returns extend the move)}} \\
1 day & 5.3 & $-$0.71 & $-$0.31 & 0.03 & 0.44 & $-$8.29 & $-$0.82 & $-$0.26 & 0.04 & 0.32 & $-$5.90 \\
1 week & 5.3 & $-$0.77 & $-$0.61 & 0.08 & -- & $-$7.13 & $-$0.48 & $-$0.56 & 0.10 & -- & $-$5.84 \\
1 month & 5.2 & $-$3.02 & $-$0.78 & 0.21 & 0.26 & $-$3.23 & $-$2.72 & $-$0.77 & 0.25 & 0.28 & $-$2.73 \\
1 quarter & 5.2 & $-$7.05 & $-$1.59 & 0.41 & 0.23 & $-$3.35 & $-$6.32 & $-$1.21 & 0.62 & 0.19 & $-$1.65 \\
\addlinespace[3pt]\multicolumn{12}{@{}l}{\textit{Panel B. Growth of sales per share, all event days}} \\
1 month & 5.5 & 3.99 & 1.85 & 0.21 & 0.46 & 7.79 & 3.63 & 1.61 & 0.21 & 0.44 & 7.00 \\
1 quarter & 5.5 & 7.48 & 2.65 & 0.33 & 0.35 & 6.44 & 7.35 & 2.48 & 0.34 & 0.34 & 5.89 \\
1 year & 5.2 & 26.53 & 7.56 & 0.73 & 0.28 & 7.99 & 28.30 & 7.94 & 1.03 & 0.28 & 5.94 \\
2 years & 4.8 & 44.01 & 6.58 & 1.18 & 0.15 & 3.72 & 48.21 & 8.16 & 0.98 & 0.17 & 5.41 \\
\addlinespace[3pt]\multicolumn{12}{@{}l}{\textit{Panel C. Growth of sales per share, announcement days}} \\
1 month & 5.7 & 6.72 & 4.04 & 0.36 & 0.60 & 9.79 & 6.21 & 3.72 & 0.34 & 0.60 & 9.72 \\
1 quarter & 5.7 & 10.45 & 5.27 & 0.66 & 0.50 & 6.80 & 10.45 & 5.33 & 0.71 & 0.51 & 6.41 \\
1 year & 5.4 & 29.09 & 11.08 & 1.27 & 0.38 & 6.98 & 30.93 & 13.16 & 1.74 & 0.43 & 6.21 \\
2 years & 5.0 & 42.13 & 8.84 & 2.07 & 0.21 & 3.23 & 46.20 & 14.75 & 1.66 & 0.32 & 7.22 \\
\addlinespace[3pt]\multicolumn{12}{@{}l}{\textit{Panel D. Growth of sales per share, other event days}} \\
1 month & 5.3 & 1.82 & 0.10 & 0.21 & 0.06 & 0.04 & 1.64 & $-$0.02 & 0.22 & $-$0.01 & $-$0.44 \\
1 quarter & 5.3 & 5.12 & 0.58 & 0.33 & 0.11 & 0.92 & 4.96 & 0.28 & 0.26 & 0.06 & 0.18 \\
1 year & 5.0 & 24.50 & 4.76 & 0.94 & 0.19 & 3.67 & 26.26 & 3.91 & 0.97 & 0.15 & 2.66 \\
2 years & 4.6 & 45.53 & 4.76 & 1.45 & 0.10 & 1.64 & 49.79 & 2.96 & 1.16 & 0.06 & 0.42 \\
\bottomrule
\end{tabular}}
\floatnote{The table reports the contribution of the text part of the event-day move to the whole move's covariance with later returns and with the later growth of sales per share, in percent of the variance of the whole move (the whole unexpected event-day return). ``Text Share'' is that of the readout of the event return used at the horizon. ``Whole Move'' is the covariance of the whole move with the unexpected later outcome, ``Text Part'' the contribution of the text part to it, with its standard error (SE), and ``Fraction'' their ratio. The $z$-statistic tests that the text part contributes in proportion to its variance. In Panel A a dash replaces the one-week fraction, because the whole move's covariance with one-week returns is within two standard errors of zero. Later returns enter with the sign reversed. In the chronological fit, whose readout of the event return needs no gap, the text share is the same at every horizon: \N{ctsA}\%, \N{ctsB}\%, \N{ctsC}\%, and \N{ctsD}\% in Panels A to D. Panel A uses \N{nA} firm-days (292,427 in the chronological fit) and Panels B to D the \N{nVS} with sales (271,526), of which \N{nVSea} (77,532) are announcement days. Standard errors allow for firm clusters and for common shocks across firms on the same and nearby days.}
\end{table}

\subsection{Extended by Later Returns}

Two known facts serve as benchmarks. Price moves on days without news are partly reversed \citep{chan2003stock,savor2012stock}: on such days later returns take back \N{nnd}\% of the move after a day, \N{nnw}\% after a week, and \N{nnq}\% after a quarter (Table~\ref{tab:levels}). Price moves on news days are not reversed \citep{tetlock2010does}, and after earnings announcements they are extended \citep{bernard1989post}. On event days the offset share of the whole move is \N{osWLw} after a week, with a standard error of \N{osWLsew}, \N{osWLm} after a month, and \N{osWLq} after a quarter.

The accounting shows that, in the first week, the extension of an event-day move comes disproportionately from the text part. In the baseline fit the offset share of the text part is \N{osTd} after a day and \N{osTw} after a week, against \N{osUd} and \N{osUw} for the remainder; the differences are \N{osZd} and \N{osZw} standard errors from zero. Later returns thus add about a tenth to the text part of a move within a week, half of it on the first day, and they add almost nothing to the remainder. The contributions show the same pattern. In Panel A of Table~\ref{tab:contrib}, the one-week extension of the whole move is \N{kWw} percentage points of its variance. The text part, which is \N{kTsw}\% of that variance, contributes \N{kTw} of these points, and the hypothesis that it contributes in proportion to its variance is rejected ($|z|=\N{kZabsw}$).

The chronological fit gives the same picture. For 2017--2023 the offset share of the text part after a week is \N{cosTw}, against \N{cosUw} for the remainder (Table~\ref{tab:week}), and its contribution to the extension is \N{ckTw} points with a standard error of \N{ckTsew} ($|z|=\N{ckZabsw}$).

The extension is larger in point estimate at a month and a quarter. In the baseline fit the offset share of the text part is \N{osTm} after a month and \N{osTq} after a quarter, against \N{osUm} and \N{osUq} for the remainder (Table~\ref{tab:levels}). The contributions of the text part differ from proportional ones by \N{kZAm} and \N{kZAq} standard errors in the baseline fit and by \N{ckZAm} and \N{ckZAq} in the chronological fit. The evidence is thus strongest in the first week; at a month it holds in both fits, and at a quarter in the baseline fit. Beyond a quarter the offset share of the text part is not estimated precisely enough to compare with that of the remainder (Table~\ref{tab:long}). The remainder's offset share is \N{lgRestY} after a year and \N{lgRestYY} after two, so event-day moves as a whole are extended at long horizons in this sample, a pattern that is not specific to the text.

Where the extension occurs is informative (Table~\ref{tab:week}). It is strongest in the smallest third of firms, where the one-week offset share of the text part is \N{osSmall} in the baseline fit and \N{cosSmall} in the chronological fit, and about half as large in the middle third. In the largest third it is \N{osBig} and \N{cosBig}, \N{osBigZ} and \N{cosBigZ} standard errors from the remainder. The extension is present on announcement days (\N{osEa}) and on other event days (\N{osNoEa}), \N{osEaZ} and \N{osNoEaZ} standard errors from the remainder, so it reaches beyond the drift after earnings announcements of \citet{bernard1989post}, which \citet{meursault2023pead} relate to the text of earnings calls. It is no smaller than in the full sample for filings accepted after the close (\N{osPost}), whose event day contains a full trading session after the text is public, and on the \N{nofollowShare}\% of firm-days that are followed by no further current report or earnings announcement by the firm within five days (\N{osNoFollow}).

\subsubsection{A Sort That Could Have Been Made at the Time}

The estimates above are covariances and give weight to large moves. Table~\ref{tab:sort} asks the same question in the way an investor would. On each event day in 2017--2023 we take the text part from the chronological fit, which uses nothing dated after the close of that day, and sort firm-days into fifths with breakpoints from the three preceding months. Returns in this table are simple returns relative to the market, the kind an investor earns; the rest of the paper uses log returns, which the identity requires.

\begin{table}[tbp]
\caption{Returns after the Event Day, by Fifth of the Event-Day Signal}\label{tab:sort}
\tabwrap{\setlength{\tabcolsep}{4pt}\begin{tabular}{@{}lrrrrrrrr@{}}
\toprule
 & Lowest & & & & Highest & Highest & & \\
Sample and Signal & Fifth & 2 & 3 & 4 & Fifth & Minus Lowest & SE & $t$ \\
\midrule
\multicolumn{9}{@{}l}{\textit{Panel A. Return over the following week}} \\
All firm-days: text part & $-$48.3 & $-$28.2 & $-$11.2 & $-$7.5 & $-$9.5 & 38.9 & 11.1 & 3.49 \\
\quad Remainder & $-$44.4 & $-$13.1 & $-$5.3 & $-$11.1 & $-$30.7 & 13.7 & 10.8 & 1.26 \\
\quad Whole move & $-$51.7 & $-$9.4 & $-$6.3 & $-$14.3 & $-$23.1 & 28.5 & 11.3 & 2.53 \\
Smallest third: text part & $-$75.6 & $-$60.7 & $-$8.2 & $-$15.8 & $-$19.4 & 56.2 & 15.0 & 3.74 \\
\quad Remainder & $-$86.4 & $-$25.4 & $-$0.2 & $-$16.8 & $-$45.2 & 41.2 & 18.7 & 2.21 \\
\quad Whole move & $-$101.5 & $-$8.4 & $-$10.0 & $-$26.6 & $-$32.1 & 69.4 & 20.0 & 3.46 \\
Middle third: text part & $-$34.5 & $-$20.9 & $-$14.0 & 1.4 & $-$4.4 & 30.1 & 13.6 & 2.21 \\
\quad Remainder & $-$10.7 & $-$12.7 & $-$7.8 & $-$13.8 & $-$27.8 & $-$17.1 & 14.0 & $-$1.22 \\
\quad Whole move & $-$16.5 & $-$11.3 & $-$8.4 & $-$12.6 & $-$24.1 & $-$7.6 & 14.0 & $-$0.55 \\
Largest third: text part & 2.7 & $-$13.2 & $-$10.4 & $-$7.7 & 3.0 & 0.3 & 10.9 & 0.02 \\
\quad Remainder & $-$18.6 & $-$3.7 & $-$6.5 & $-$4.3 & $-$9.2 & 9.4 & 9.7 & 0.97 \\
\quad Whole move & $-$18.0 & $-$8.6 & $-$2.5 & $-$5.5 & $-$7.1 & 10.9 & 10.0 & 1.09 \\
Announcement days: text part & $-$49.1 & $-$20.2 & 0.0 & $-$11.1 & 16.4 & 65.5 & 14.7 & 4.46 \\
\quad Remainder & $-$33.1 & $-$22.2 & $-$1.5 & $-$6.7 & $-$3.4 & 29.7 & 11.7 & 2.54 \\
\quad Whole move & $-$45.4 & $-$16.8 & 5.4 & $-$10.3 & 5.7 & 51.2 & 11.6 & 4.43 \\
Other event days: text part & $-$47.7 & $-$29.9 & $-$12.4 & $-$6.4 & $-$29.4 & 18.3 & 14.5 & 1.26 \\
\quad Remainder & $-$52.4 & $-$11.1 & $-$5.9 & $-$12.3 & $-$51.9 & 0.5 & 15.7 & 0.03 \\
\quad Whole move & $-$56.2 & $-$7.9 & $-$8.2 & $-$15.3 & $-$46.1 & 10.1 & 16.6 & 0.61 \\
\addlinespace[3pt]\multicolumn{9}{@{}l}{\textit{Panel B. Return over the following month}} \\
All firm-days: text part & $-$131.4 & $-$87.8 & $-$67.9 & $-$52.1 & $-$64.2 & 67.2 & 26.2 & 2.56 \\
\quad Remainder & $-$143.0 & $-$64.9 & $-$50.1 & $-$52.9 & $-$92.3 & 50.7 & 19.6 & 2.58 \\
\quad Whole move & $-$153.3 & $-$54.8 & $-$47.0 & $-$66.2 & $-$82.1 & 71.2 & 19.1 & 3.72 \\
\bottomrule
\end{tabular}}
\floatnote{The table reports average returns relative to the market, in basis points, from the close of the event day over the following 5 or 21 trading days: the gross return of the stock divided by the gross return of the value-weighted market, less one. Firm-days are sorted on a signal known at that close: the text part of the event-day move from the chronological fit, the remainder of the move, or the whole unexpected move. Breakpoints between fifths come from the three preceding calendar months. The sample is the 292,427 firm-days of 2017--2023. No trading costs are deducted. SE and $t$ are the standard error and the $t$-statistic of the difference between the highest and lowest fifths; standard errors allow for firm clusters and for common shocks across firms on the same and nearby days.}
\end{table}

Firm-days in the highest fifth of the text part outperform those in the lowest by \N{sTw} basis points over the following week ($t=\N{sTwt}$). A sort on the remainder gives \N{sUw} basis points ($t=\N{sUwt}$), although the remainder is by far the larger part of the move, and a sort on the whole move gives \N{sWw}. The spread from the text part is \N{sTwsmall} basis points in the smallest third of firms ($t=\N{sTwsmallt}$), \N{sTwmid} in the middle third ($t=\N{sTwmidt}$), and \N{sTwbig} in the largest ($t=\N{sTwbigt}$). It is \N{sTwea} basis points on announcement days ($t=\N{sTweat}$) and \N{sTwnoea} on other event days ($t=\N{sTwnoeat}$). In log returns the spread for all firm-days is \N{sLTw} basis points ($t=\N{sLTwt}$; untabulated). The spread comes mainly from the low returns of the two lowest fifths. For all firm-days every fifth has a negative average: the benchmark is the value-weighted market and the sample gives every firm-day the same weight, so the averages reflect the returns of small firms relative to large ones in 2017--2023; in the largest third they are closer to zero and in places positive.

The sort and the covariances weight firm-days differently and do not agree in every subsample. On other event days the spread from the text part is not distinguishable from zero, although its offset share in the same fit is \N{cosNoEaZ} standard errors from that of the remainder (Table~\ref{tab:week}). In the smallest third the remainder also sorts later returns, by \N{sUwsmall} basis points ($t=\N{sUwsmallt}$), and the whole move gives the widest spread there, \N{sWwsmall}. On announcement days the remainder sorts later returns by \N{sUwea} basis points ($t=\N{sUweat}$), although its offset share is near zero. Over a month the sorts on the text part, the remainder, and the whole move give spreads of \N{sTm}, \N{sUm}, and \N{sWm} basis points. The spreads are before trading costs and are largest in small stocks, where trading is costly.

\subsubsection{Possible Readings}

The example of Section~\ref{sec:example} offers a reading of each of the three kinds of move. A positive offset share is what short-lived discount-rate news or an overreaction would produce, as would price pressure; moves on days without news show it. An offset share of zero corresponds to a complete reaction, and it is what the covariances show for the remainder of an event-day move during the first week. A negative offset share corresponds to a reaction that is too small: on that reading the market underreacts to the part of a filing that our readout captures and reacts within the day to most of the rest of the day's news. Three other mechanisms would also extend a move: a change in required returns after a filing; slow trading in illiquid stocks, which fits the strength of the effect in small firms; and news from other sources that arrives in the following days and is correlated with the filing. The firm's own later current reports and earnings announcements are not the source: the extension is no smaller on the firm-days that are followed by neither within five days. Whichever mechanism is at work, the accounting separates two parts of the same day's price move that are followed by different returns.

\subsection{Realized in Fundamentals}

Every kind of price move is followed by growth in fundamentals. A year after an event day, \N{gAWry}\% of the whole move has been matched by growth in sales per share, and after two years \N{gAWryy}\%; for moves on days without news the figures are \N{nnGy}\% and \N{nnGyy}\% (Table~\ref{tab:levels}). This is the relation between returns and later fundamentals that \citet{kothari1992information} and \citet{collins1994lack} document for earnings, expressed here for sales and as a share of the price move.

The text part is a small part of the price move and a large part of its relation to later sales: per unit of variance, it is followed by far more sales growth than the remainder (Panel B of Table~\ref{tab:contrib}). After a month it contributes \N{gATm} of the \N{gAWm} percentage points of the whole move's growth share, or \N{gAFm}\%, from \N{gATsm}\% of the variance. After a year it contributes \N{gATy} of \N{gAWy} points, or \N{gAFy}\%, with a 95\% confidence set of \N{fielLo}\% to \N{fielHi}\% by the method of \citet{fieller1954some}, and after two years \N{gAFyy}\%. The $z$-statistics against a proportional contribution are between \N{zminA} and \N{zmaxA}. The chronological fit gives fractions of \N{cgAFm}\%, \N{cgAFq}\%, \N{cgAFy}\%, and \N{cgAFyy}\% at the four horizons, with $z$-statistics of \N{cgAZrm}, \N{cgAZrq}, \N{cgAZry}, and \N{cgAZryy}. The estimated growth share of the text part exceeds one at one and two years (Table~\ref{tab:levels}): the displacement described above raises these levels, and the contributions are free of it.

After a month, less of the text part than of the remainder is still in the valuation ratio: its valuation share is \N{vsTm}, against \N{vsUm} for the remainder, a difference of \N{vsZm} standard errors (Table~\ref{tab:levels}). After a quarter the two shares are \N{vsTq} and \N{vsUq} and differ by \N{vsZq} standard errors, and beyond a quarter the standard error of the valuation share of the text part exceeds 0.6.

\subsection{Results Reported and Sales Earned Later}

The example of Section~\ref{sec:example} separates news about cash flows of the current period, $\varepsilon^{w}$, which enters the fundamental at once, from news about growth, $\eta^{w}$, which is realized at the rate at which the growth arrives. The timing of financial reports allows a related separation in the data, between sales that are reported after a filing and sales that are earned after it. Panels C and D of Table~\ref{tab:contrib} separate announcement days from other event days. On announcement days the text part contributes \N{gEFm}\% of the growth share after a month. On other event days it contributes no more than in proportion to its variance within a quarter, and \N{gNFy}\% after a year ($z=\N{gNZy}$); after two years the fraction is \N{gNFyy}\% and its difference from a proportional contribution is \N{gNZyy} standard errors. These totals mix sales of different vintages, and the mix matters for what the result means.

\begin{table}[tbp]
\caption{Growth of Sales per Share by When the Sales Were Earned}\label{tab:vintage}
\tabwrap{\setlength{\tabcolsep}{4pt}\begin{tabular}{@{}lrrrrrrrr@{}}
\toprule
 & \multicolumn{4}{c}{Announcement Days} & \multicolumn{4}{c}{Other Event Days} \\
\cmidrule(lr){2-5}\cmidrule(l){6-9}
 & Whole & Text & & & Whole & Text & & \\
Quarters, by When Earned & Move & Part & SE & $z$ & Move & Part & SE & $z$ \\
\midrule
\multicolumn{9}{@{}l}{\textit{Panel A. Baseline, 2014--2023: 1 year after the event day}} \\
\quad Announced on the event day & 6.75 & 4.77 & 0.47 & 8.83 & $-$0.16 & 0.18 & 0.06 & 3.20 \\
\quad Ended before the event day, reported later & 0.46 & $-$0.18 & 0.09 & $-$2.40 & $-$0.97 & $-$0.29 & 0.22 & $-$1.10 \\
\quad In progress on the event day & 7.01 & 3.30 & 0.51 & 5.48 & 4.25 & 2.20 & 0.33 & 6.24 \\
\quad Begun after the event day & 13.37 & 4.60 & 0.75 & 5.09 & 10.49 & 3.58 & 0.67 & 4.73 \\
\quad Change in the number of shares & 1.50 & $-$1.31 & 0.82 & $-$1.71 & 10.89 & $-$1.04 & 0.49 & $-$3.18 \\
Total: growth of sales per share & 29.09 & 11.08 & 1.27 & 6.98 & 24.50 & 4.76 & 0.94 & 3.67 \\
\addlinespace[3pt]\multicolumn{9}{@{}l}{\textit{Panel B. Baseline, 2014--2023: 2 years after the event day}} \\
\quad Announced on the event day & 6.81 & 4.54 & 0.38 & 10.17 & $-$0.16 & 0.17 & 0.06 & 2.97 \\
\quad Ended before the event day, reported later & 0.51 & $-$0.04 & 0.13 & $-$0.54 & $-$0.86 & $-$0.14 & 0.19 & $-$0.52 \\
\quad In progress on the event day & 7.02 & 3.37 & 0.52 & 5.48 & 4.58 & 2.20 & 0.35 & 6.05 \\
\quad Begun after the event day & 21.34 & 4.35 & 2.03 & 1.64 & 19.27 & 4.47 & 1.05 & 3.47 \\
\quad Change in the number of shares & 6.44 & $-$3.12 & 1.67 & $-$2.13 & 22.70 & $-$1.94 & 0.96 & $-$3.09 \\
Total: growth of sales per share & 42.13 & 8.84 & 2.07 & 3.23 & 45.53 & 4.76 & 1.45 & 1.64 \\
\addlinespace[3pt]\multicolumn{9}{@{}l}{\textit{Panel C. Chronological, 2017--2023: 1 year after the event day}} \\
\quad Announced on the event day & 6.48 & 4.10 & 0.28 & 11.69 & $-$0.22 & 0.16 & 0.09 & 2.03 \\
\quad Ended before the event day, reported later & 0.41 & 0.06 & 0.13 & 0.33 & $-$0.71 & $-$0.30 & 0.16 & $-$1.88 \\
\quad In progress on the event day & 7.09 & 3.25 & 0.61 & 4.51 & 4.20 & 2.16 & 0.37 & 5.88 \\
\quad Begun after the event day & 14.22 & 5.57 & 0.91 & 5.22 & 13.14 & 3.67 & 0.59 & 5.20 \\
\quad Change in the number of shares & 2.73 & 0.36 & 0.36 & 0.63 & 9.84 & $-$2.05 & 0.35 & $-$6.70 \\
Total: growth of sales per share & 30.93 & 13.16 & 1.74 & 6.21 & 26.26 & 3.91 & 0.97 & 2.66 \\
\bottomrule
\end{tabular}}
\floatnote{The table splits the growth of sales per share after the event day into the parts due to fiscal quarters, grouped by when their sales were earned, and the part due to the change in the number of shares, with the sign reversed. Entries other than $z$ are in percent of the variance of the whole unexpected event-day return: the covariance of the whole move, and of its text part, with each piece. The whole-move pieces sum to the total; the text-part pieces do so approximately, because each piece has its own fitted readout. SE is the standard error of the text part's contribution, and $z$ tests that the text part contributes in proportion to its variance. For other event days the first row is not empty because Compustat dates some releases to an event day on which the Institutional Brokers' Estimate System (I/B/E/S) records no announcement. Samples are those of Table~\ref{tab:contrib}.}
\end{table}

A later accounting report can contain sales that were earned before the filing appeared. Table~\ref{tab:vintage} therefore splits the growth of sales per share into the five pieces of Section~\ref{sec:outcomes}. The timing can be checked directly (the counts in this paragraph are untabulated). Under our dating rule, the quarter announced on an announcement day is the only quarter to enter trailing sales during the following month in 98\% of cases. After a year, an average of 4.5 quarters have entered: 0.98 announced on the event day, 0.98 in progress on it, and 2.58 that began after it. On other event days, 3.8 quarters have entered, of which 0.35 had ended before the event day, 0.96 were in progress, and 2.5 began after it.

On announcement days, the text part is most closely tied to the quarter being announced. A year after the event, that quarter accounts for \N{vESETy} of the \N{gETy} points that the text part contributes, and the text part carries \N{vESEFy}\% of the whole move's comovement with it. Within an announcement return, then, the part that a reading of the filing accounts for is closely related to the results just reported, and the remainder is much less so. The quarter in progress adds \N{vESPTy} points, or \N{vESPFy}\% of the whole move's comovement with that quarter, and the quarters that began after the announcement add \N{vESATy} points, or \N{vESAFy}\%, with a standard error of \N{vESATsey}.

On other event days I/B/E/S records no announcement, and the pattern differs. The text part shows no positive relation to sales that had been earned but not yet reported: its contribution through quarters that had ended before the event day is \N{vNSBTy} points, with a standard error of \N{vNSBTsey}. It is related to the quarter in progress, where it contributes \N{vNSPTy} of \N{vNSPWy} points, and to the quarters that began after the event day, where it contributes \N{vNSATy} of \N{vNSAWy} points, or \N{vNSAFy}\%, from \N{gNTsy}\% of the variance ($z=\N{vNSAZy}$). After two years the contribution of those later quarters is \N{vNSATyy} of \N{vNSAWyy} points ($z=\N{vNSAZyy}$). In the chronological fit for 2017--2023, they contribute \N{cvNSATy} points after a year, with a standard error of \N{cvNSATsey} ($z=\N{cvNSAZy}$), and quarters that had ended before the event day contribute \N{cvNSBTy}. On other event days, then, the part of a price move that the text accounts for is followed by the sales of the quarter in progress and of later quarters, not by the sales of quarters that had ended before the filing and were reported after it. In the example this is how news about growth would appear.

The number of shares moves the other way. On other event days, the text part's contribution through the share count is \N{vNNSTy} points after a year, with a standard error of \N{vNNSTsey}: favorable text is followed by more shares, which lowers sales per share. The growth of total sales that follows the text part is therefore larger than the growth per share.

What sales growth implies for the cash flows of shareholders depends on margins and on the investment needed to produce the sales. Table~\ref{tab:fund} therefore repeats the accounting for two other fundamentals. With book equity per share, a balance-sheet item that each quarter's earnings enter once, the text part carries \N{bkAFm}\% of the growth share after a month ($z=\N{bkAZm}$) and \N{bkAFy}\% after a year, in line with its share of the variance ($z=\N{bkAZy}$). For the change in trailing operating income, scaled by market value before the event, an outcome outside the identity, it contributes \N{oiAT} of the \N{oiAW} points of the whole move's comovement over a year, or \N{oiAF}\% ($z=\N{oiAZ}$), and \N{oiNT} of \N{oiNW} points on other event days ($z=\N{oiNZ}$). The relation of the text part to later sales thus extends to operating income over a year.

\subsection{Robustness and the Role of the Readout}\label{sec:robust}

Table~\ref{tab:readouts} shows how the two main contributions vary with the readout and with the way it is fitted. The gap matters little for these two contributions: without it they are \N{rNoGapWk} and \N{rNoGapG} points. It matters at long horizons. Without the gap the text part appears to be extended for years, with an offset share of \N{lgNoGapyyy} after three years (standard error \N{lgNoGapSeyyy}); with the gap the estimate is \N{lgGapyyy} (standard error \N{lgGapSeyyy}; Table~\ref{tab:long}).

\begin{table}[tbp]
\caption{The Main Contributions under Different Readouts and Fitting Rules}\label{tab:readouts}
\tabwrap{\begin{tabular}{@{}lrrrrrrr@{}}
\toprule
 & Text & \multicolumn{3}{c}{Later Returns, 1 Week} & \multicolumn{3}{c}{Sales Growth, 1 Year} \\
\cmidrule(lr){3-5}\cmidrule(l){6-8}
Readout & Share (\%) & Text Part & SE & $z$ & Text Part & SE & $z$ \\
\midrule
\multicolumn{8}{@{}l}{\textit{Panel A. 2014--2023}} \\
Baseline: cross-fitted with the gap & 5.49 & $-$0.61 & 0.08 & $-$7.13 & 7.56 & 0.73 & 7.99 \\
Without the gap & 5.49 & $-$0.61 & 0.08 & $-$7.50 & 6.97 & 0.58 & 9.36 \\
Without the group scales & 4.15 & $-$0.47 & 0.06 & $-$7.49 & 6.12 & 0.66 & 7.28 \\
With 1,024 nonlinear features & 5.62 & $-$0.62 & 0.08 & $-$7.26 & 7.61 & 0.71 & 8.34 \\
With the volatility adjustment & 4.66 & $-$0.44 & 0.11 & $-$3.85 & 6.71 & 0.70 & 7.64 \\
\addlinespace[3pt]\multicolumn{8}{@{}l}{\textit{Panel B. 2017--2023}} \\
Baseline: cross-fitted with the gap & 5.35 & $-$0.60 & 0.09 & $-$6.48 & 7.66 & 0.91 & 6.46 \\
Chronological & 4.71 & $-$0.56 & 0.10 & $-$5.84 & 7.94 & 1.03 & 5.94 \\
Chronological, without the group scales & 3.66 & $-$0.41 & 0.08 & $-$4.92 & 5.96 & 0.96 & 4.92 \\
Chronological, with the volatility adjustment & 3.46 & $-$0.29 & 0.14 & $-$2.08 & 9.82 & 2.94 & 2.75 \\
\bottomrule
\end{tabular}}
\floatnote{The table reports, for variants of the readout, the text share of the event return and the contribution of the text part to the one-week offset and to one-year sales growth, in percent of the variance of the whole unexpected event-day return as in Table~\ref{tab:contrib}; $z$ tests that a contribution is in proportion to the text part's variance. Text shares are for the readout of the event return fitted without a gap. ``Without the gap'' fits the readout of the event return on all quarters other than the one predicted. ``Without the group scales'' omits the nine scale factors. ``With 1,024 nonlinear features'' appends random nonlinear features of the embedding, so that the class of readouts contains the linear one. ``With the volatility adjustment'' makes the response proportional to the firm's volatility over the previous 63 days. ``Chronological'' rows fit every step on earlier months only. SE is the standard error of the contribution.}
\end{table}

A more flexible class of readouts changes the results little: appending 1,024 nonlinear features to the embedding gives a text share of \N{tsRF}\% and contributions of \N{rRFWk} and \N{rRFG} points. Without the group scales the text share is \N{tsNoGroups}\% and the contributions are \N{rNoGroupsWk} and \N{rNoGroupsG} points. The chronological fit without the group scales, a ridge readout that is linear in the embedding with no further adjustment, gives \N{chNoScWk} points ($z=\N{chNoScWkZ}$) and \N{chNoScG} points ($z=\N{chNoScGZ}$). The design matters for both size and precision. With the volatility adjustment, the text share is \N{tsAlt}\% and the two contributions are \N{rAltWk} and \N{rAltG} points; in the chronological fit the one-week contribution is then \N{rAltChWk} points ($z=\N{rAltChWkZ}$), against \N{rChWk} ($z=\N{rChWkZ}$) for the design we use, and the contribution to one-year sales growth is \N{rAltChG} points ($z=\N{rAltChGZ}$), against \N{rChG} ($z=\N{rChGZ}$). In every row of Table~\ref{tab:readouts} the two contributions keep their sign and differ from proportional ones by more than two standard errors.

Several checks bear on the inference (Tables~\ref{tab:boot} to~\ref{tab:placebo} and~\ref{tab:week}). The bootstrap, which resamples blocks of calendar quarters with the fitted readouts held fixed, gives a $z$-statistic of \N{kZbootw} for the one-week contribution, against \N{kZonew} with the analytic standard error, and \N{gAZbooty} against \N{gAZry} for one-year sales growth; for the one-month result on announcement days it gives \N{gEZbootm} against \N{gEZrm}. Under estimators of the variance with fewer or more lags and under a jackknife over calendar quarters, no test statistic of the main contributions falls by more than 0.10 in absolute value (Table~\ref{tab:sesens}). When embeddings are reassigned at random across firms within each day, the two contributions are \N{plWk} and \N{plG} points; the one-week figure is \N{plWkZ} standard errors from zero and less than a hundredth of the estimate. The one-week result holds in each part of the decade: the offset share of the text part is \N{osPerA}, \N{osPerB}, and \N{osPerC} in 2014--2016, 2017--2019, and 2020--2023.

Log returns have long tails in this sample (Table~\ref{tab:sumstats}), so we also compute the contributions with each later outcome winsorized at its 1st and 99th percentiles. The one-week contribution is then \N{wWk} points, against \N{wWkRaw} before winsorizing with the same simple estimator (the covariance without the correction term of the orthogonal score), and the contribution to one-year sales growth is \N{wG} points, against \N{wGRaw} (Table~\ref{tab:winsor}).

The two contributions are stable across quarters, firms, and years (Table~\ref{tab:stab}). In the baseline fit, when any one calendar quarter is left out, the one-week contribution stays between $-$\N{loqWkHi} and $-$\N{loqWkLo} points and the contribution to one-year sales growth between \N{loqGLo} and \N{loqGHi}; leaving out any one firm moves them less. The difference from a proportional contribution has its full-sample sign in \N{yrsWk} calendar years for later returns and in \N{yrsG} for sales, and in every year of the chronological fit. For the one-year horizon, twelve subsamples, each made of the event days of one calendar month in every year, have outcome windows that overlap by no more than about a month from one year to the next, and all twelve give a contribution to sales growth above the proportional one. The contributions are about the same or larger without the firm-days whose stock is delisted within the horizon and without those with the most extreme returns and growth rates.

\section{Conclusion}\label{sec:conclusion}

Language models can read every filing a firm makes, but their output has no economic units. We define the meaning of a text by the expectations it revises, and we choose the outcomes so that the revisions add up to the part of the return that the text accounts for. The definition gives the content of a text economic units, and it yields a decomposition of an event-day price move that rests on an exact identity and in which no term is a residual.

Applied to a decade of current reports, the decomposition shows that the part of an event-day move that the text accounts for, about one-twentieth of its variance, has a distinct economic signature. Later returns extend it by about a tenth within a week, mainly in smaller firms, while they add almost nothing to the remainder of the same move and partly reverse moves on days without news. Later sales growth matches far more of it than of the remainder: it carries more than a quarter of the comovement between event-day returns and the growth of sales over the following year. On announcement days it is most closely tied to the results just announced, and on other event days it is followed by the sales of the quarter in progress and of quarters that began after the filing, and not by the sales of quarters that had already ended. The estimates are similar in size in the chronological fit.

These patterns separate the text part of a price move from the remainder, whichever mechanism produces them. In the model the one-week pattern is that of too small a reaction to the content of filings, and the sales pattern is that of news about future growth; data on expectations and on trading would distinguish these readings from changes in required returns and from slow trading. Richer readouts and longer samples will extend the accounting beyond the linear component of a text and beyond a single decade. The construction itself is general. Wherever an identity links outcomes, anchoring an embedding to those outcomes turns it into quantities that are measured in economic units and that add up, so that the content of a text can be accounted for as well as used for prediction.

\section*{Data and Code Availability}
The current reports are public and were downloaded from the EDGAR system of the Securities and Exchange Commission. Returns, accounting data, analyst data, option data, and institutional holdings are from CRSP, Compustat, I/B/E/S, OptionMetrics, and Thomson Reuters, obtained through Wharton Research Data Services under license, and cannot be redistributed by the authors. The language model is publicly available. The code is available from the corresponding author on request.

\clearpage
\appendix
\appendixformat
\section{Variable Definitions and Summary Statistics}\label{app:vars}
Table~\ref{tab:defs} defines the outcomes of the accounting and the terms used for the components of a price move. Table~\ref{tab:sumstats} reports summary statistics for the two estimation samples.

\begin{table}[H]
\caption{Variable Definitions}\label{tab:defs}
\tabwrap{\begin{tabular}{@{}p{0.27\textwidth}p{0.69\textwidth}@{}}
\toprule
Variable & Definition \\
\midrule
Event return, $r^{E}$ & Log return, including dividends and delisting returns, from the close of day $d-1$ to the close of day $d$ (CRSP). Market adjusted: net of the log return on the value-weighted market. \\[3pt]
Later return, $R_{H}$ & Log return from the close of day $d$ to the close of day $d+H$, market adjusted. \\[3pt]
Growth of the fundamental, $G_{H}$ & Change in the log of the fundamental per share from the value known at the close of day $d-1$ to the value known at the close of day $d+H$. The fundamental is trailing four-quarter sales or book equity (Compustat); a quarter is known from the first close after the release of its results. \\[3pt]
Payout, $D_{H}$ & Sum of log dividend yields from day $d$ to day $d+H$: the cumulative log return less the cumulative log price change. \\[3pt]
Change in valuation, $V_{H}$ & Change in the log ratio of market capitalization to the fundamental from the close of day $d-1$ to the close of day $d+H$, net of the cumulative log return on the market. \\[3pt]
Text part, $c^{r}$ & Fitted revision in the expected event return due to the text, given the state at the close of day $d-1$ (Definition~\ref{def:concept}); cross-fitted. \\[3pt]
Remainder, $u^{r}$; whole move, $\tilde r^{E}$ & The whole move is the event return less its expectation given the state. The remainder is the whole move less the text part. \\[3pt]
Text share & Variance of the text part divided by the variance of the whole move. \\[3pt]
Share, $\omega(a)$ & Covariance of a component $a$ of the whole move with the unexpected part of a later outcome, divided by the variance of $a$, with the sign reversed for later returns. \\[3pt]
Contribution & Covariance of the text part with the unexpected part of a later outcome, divided by the variance of the whole move, with the sign reversed for later returns. \\[3pt]
Announcement day & Event day on which I/B/E/S records an earnings announcement by the firm. \\
\bottomrule
\end{tabular}}
\floatnote{The table defines the outcomes that enter the identity of Proposition~\ref{prop:fates} and the components of the event-day move. Day $d$ is the event day, the first trading day whose close follows the acceptance of the filing, and $H$ is the horizon in trading days. All returns and growth rates are log changes. CRSP is the Center for Research in Security Prices, and I/B/E/S is the Institutional Brokers' Estimate System.}
\end{table}

\begin{table}[H]
\caption{Summary Statistics}\label{tab:sumstats}
\tabwrap{\setlength{\tabcolsep}{2.8pt}\begin{tabular}{@{}lrrrrrrrr@{}}
\toprule
Variable & $N$ & Mean & Std.\ Dev. & Min. & P1 & Median & P99 & Max. \\
\midrule
\multicolumn{9}{@{}l}{\textit{Panel A. Returns sample}} \\
Event return, raw & 406,233 & $-$0.20 & 8.03 & $-$289.74 & $-$24.05 & 0.00 & 20.01 & 211.71 \\
Event return, market adjusted & 406,233 & $-$0.25 & 7.93 & $-$290.78 & $-$23.82 & $-$0.05 & 19.69 & 212.91 \\
\quad Previous close to open (raw) & 406,233 & 0.05 & 6.08 & $-$272.94 & $-$16.73 & 0.01 & 14.95 & 255.39 \\
\quad Open to close (raw) & 406,233 & $-$0.24 & 5.36 & $-$214.07 & $-$16.32 & 0.00 & 14.04 & 178.62 \\
Later return, 1 day & 406,233 & $-$0.14 & 4.34 & $-$290.78 & $-$12.02 & $-$0.09 & 11.50 & 208.05 \\
Later return, 1 week & 406,233 & $-$0.55 & 8.84 & $-$338.60 & $-$26.35 & $-$0.30 & 23.15 & 262.69 \\
Later return, 1 month & 406,233 & $-$2.00 & 16.97 & $-$454.17 & $-$56.17 & $-$0.94 & 41.12 & 300.38 \\
Later return, 1 quarter & 406,233 & $-$5.98 & 29.71 & $-$862.54 & $-$108.62 & $-$2.79 & 64.21 & 294.33 \\
Later return, 1 year & 406,233 & $-$23.43 & 69.42 & $-$1,270.47 & $-$280.03 & $-$10.17 & 105.32 & 388.87 \\
Later return, 2 years & 406,102 & $-$46.14 & 110.02 & $-$3,221.39 & $-$465.85 & $-$20.49 & 124.98 & 397.70 \\
Payout, 1 year & 406,233 & 1.18 & 3.56 & 0.00 & 0.00 & 0.00 & 8.72 & 291.87 \\
Market capitalization (\$ million) & 406,233 & 9,561 & 47,736 & 0 & 10 & 884 & 177,430 & 2,988,796 \\
Stock price (\$) & 406,233 & 43.96 & 110.91 & 1.00 & 1.14 & 20.80 & 342.40 & 6,230.69 \\
Daily volatility, prior 63 days & 406,131 & 3.26 & 2.55 & 0.00 & 0.50 & 2.53 & 13.04 & 122.52 \\
Filings per firm-day & 406,233 & 1.04 & 0.22 & 1.00 & 1.00 & 1.00 & 2.00 & 12.00 \\
\addlinespace[3pt]\multicolumn{9}{@{}l}{\textit{Panel B. Sales sample}} \\
Growth of sales per share, 1 month & 382,549 & 0.25 & 19.51 & $-$921.03 & $-$44.88 & 0.00 & 36.40 & 1,198.69 \\
Growth of sales per share, 1 quarter & 382,549 & 0.71 & 31.14 & $-$935.44 & $-$87.06 & 1.27 & 72.67 & 1,198.46 \\
Growth of sales per share, 1 year & 382,549 & 2.53 & 65.48 & $-$974.45 & $-$227.30 & 5.02 & 199.05 & 1,205.85 \\
Growth of sales per share, 2 years & 382,443 & 1.79 & 93.17 & $-$2,549.89 & $-$366.09 & 8.94 & 261.79 & 1,166.55 \\
Change in valuation, 1 year & 382,549 & $-$24.45 & 83.49 & $-$1,256.31 & $-$316.72 & $-$15.85 & 185.09 & 1,086.51 \\
Change in valuation, 2 years & 382,443 & $-$44.80 & 110.61 & $-$1,420.82 & $-$441.45 & $-$30.19 & 227.57 & 1,322.63 \\
Trailing sales (\$ million) & 382,549 & 5,517 & 21,334 & 0 & 0 & 590 & 98,727 & 630,794 \\
\bottomrule
\end{tabular}}
\floatnote{The table reports summary statistics for the returns sample (\N{nA} firm-days of \N{nFirms} firms, April 2014 to December 2023) and for the firm-days with sales (\N{nVS} firm-days of \N{nVSfirms} firms). Returns, growth rates, the payout, the change in valuation, and volatility are in percent; returns and growth rates are log changes, and later returns and the change in valuation are net of the value-weighted market. Each firm-day has the same weight. $N$ is the number of firm-days, and P1 and P99 are the 1st and 99th percentiles. Daily gross returns are bounded below at 0.01 before logs are taken, a bound that does not bind in the sample. The minima and maxima lie far outside the 1st and 99th percentiles; the minimum of the one-month growth of sales per share, $-921\%$, is a fall by a factor of 10,000. Tables~\ref{tab:winsor} and~\ref{tab:stab} show that the main results do not depend on the extreme values.}
\end{table}

\clearpage
\section{Additional Results and Proofs}\label{app:proofs}
\setcounter{proposition}{0}
\renewcommand{\theproposition}{B\arabic{proposition}}
\renewcommand{\thelemma}{B\arabic{lemma}}
\makeatletter
\def\theHproposition{B\arabic{proposition}}
\def\theHlemma{B\arabic{lemma}}
\makeatother

Throughout, $L^{2}(\cH)$ denotes the square-integrable random variables measurable with respect to a $\sigma$-field $\cH$, and conditional expectation given $\cH$ is the orthogonal projection of $L^{2}$ onto $L^{2}(\cH)$. For a square-integrable random variable $Y$, define the total innovation and the text innovation,
\[
N[Y]=\E[Y\mid\Fp]-\E[Y\mid\cF],\qquad \Nw[Y]=\E[Y\mid\cG]-\E[Y\mid\cF],
\]
so that the anchored concept of outcome $k$ is $c^{k}=\Nw[Y^{k}]$.

\begin{lemma}[Text innovations]\label{lem:innov}
Let $Y$ be square integrable. Then
\begin{enumerate}[label=(\roman*)]
\item $\Nw$ is linear, and $\Nw[Y]=0$ if $Y$ is $\cF$-measurable;
\item $\E\big[\Nw[Y]\mid\cF\big]=0$;
\item if $w$ is $\cF$-measurable, then $\Nw[Y]=0$ for every $Y$.
\end{enumerate}
Under Assumption~\ref{ass:pit}, in addition,
\begin{enumerate}[label=(\roman*),resume]
\item $\Nw[Y]=\E\big[N[Y]\mid\cG\big]$;
\item $\Var(N[Y])=\Var(\Nw[Y])+\Var(N[Y]-\Nw[Y])$.
\end{enumerate}
\end{lemma}

\begin{proof}[Proof of Lemma~\ref{lem:innov}]
(i) Conditional expectation is linear, and if $Y$ is $\cF$-measurable it is also $\cG$-measurable, so both conditional expectations equal $Y$. (ii) Since $\cF\subseteq\cG$, $\E[\E[Y\mid\cG]\mid\cF]=\E[Y\mid\cF]$. (iii) In this case $\cG=\cF$. (iv) Under Assumption~\ref{ass:pit}, $\cG\subseteq\Fp$, and $\E[Y\mid\cF]$ is $\cG$-measurable, so $\E[N[Y]\mid\cG]=\E[\E[Y\mid\Fp]\mid\cG]-\E[Y\mid\cF]=\Nw[Y]$. (v) Let $U=N[Y]-\Nw[Y]=\E[Y\mid\Fp]-\E[Y\mid\cG]$. Since $\cG\subseteq\Fp$, $\E[U\mid\cG]=0$, and because $\Nw[Y]$ is $\cG$-measurable, $\E[\Nw[Y]\,U]=0$. Both terms have mean zero, so the variances add.
\end{proof}

By part (iii), stale text carries no innovation, whatever it says. Part (v) splits the variance of any piece of news into a part that the text accounts for and an orthogonal remainder.

\begin{proof}[Proof of Proposition~\ref{prop:concept}]
(i) Write $\tilde Y^{k}=Y^{k}-\E[Y^{k}\mid\cF]$. Since $\E[Y^{k}\mid\cF]\in L^{2}(\cF)\subseteq L^{2}(\cH)$, the projection of $\tilde Y^{k}$ onto $L^{2}(\cH)$ is $\E[Y^{k}\mid\cH]-\E[Y^{k}\mid\cF]=c^{k}_{\cH}$. For any $c\in L^{2}(\cH)$, $\E[(\tilde Y^{k}-c)^{2}]=\E[(\tilde Y^{k}-c^{k}_{\cH})^{2}]+\E[(c^{k}_{\cH}-c)^{2}]$, which is minimized only at $c=c^{k}_{\cH}$ almost surely. (ii) The first claim is linearity of conditional expectation. If $a^{\top}Y$ is $\cF$-measurable, then $\E[a^{\top}Y\mid\cH]=a^{\top}Y=\E[a^{\top}Y\mid\cF]$. (iii) Under the stated conditions $\sigma(\chi(x))=\sigma(x)$, so $\cH$ is unchanged.
\end{proof}

\begin{proof}[Proof of Proposition~\ref{prop:fates}]
(a) For every $s$, $r_{s}=\log(P_{s}/P_{s-1})+\log(1+\Div_{s}/P_{s})=p_{s}-p_{s-1}+\delta_{s}$. Summing from $\tau_{0}+1$ to $\tau_{1}+H$ gives $r^{E}+R_{H}=p_{\tau_{1}+H}-p_{\tau_{0}}+D_{H}$. Since $p=f+v$, $p_{\tau_{1}+H}-p_{\tau_{0}}=G_{H}+V_{H}$.

(b) Subtracting from each side of (a) its conditional expectation given $\cF$ yields $\tilde r^{E}=\tilde G_{H}+\tilde D_{H}-\tilde R_{H}+\tilde V_{H}$, and applying $\Nw$ to (a) yields the identity for concepts by Lemma~\ref{lem:innov}(i). Let $a$ be one of the three components. Taking the covariance of $a$ with both sides of the identity for unexpected parts gives $\Cov(a,\tilde r^{E})=\Var(a)\,[\omega_{G}(a)+\omega_{D}(a)+\omega_{R}(a)+\omega_{V}(a)]$. It remains to note that $\Cov(a,\tilde r^{E})=\Var(a)$ in each case: $\tilde r^{E}=c^{r}+u^{r}$, and $\E[c^{r}u^{r}]=\E[c^{r}\,\E[u^{r}\mid\cG]]=0$ because $c^{r}$ is $\cG$-measurable. Finally, for any outcome, $\Cov(c^{r},\tilde Y^{k})=\E[c^{r}\,\E[\tilde Y^{k}\mid\cG]]=\E[c^{r}c^{k}]$, which gives the expression in terms of concepts.

(c) Subtracting $\bar r^{E}+\bar R_{H}=\sum_{s=\tau_{0}+1}^{\tau_{1}+H}\bar r_{s}$ from both sides of (a) and regrouping gives $(r^{E}-\bar r^{E})=G_{H}+D_{H}-(R_{H}-\bar R_{H})+(V_{H}-\bar r^{E}-\bar R_{H})$. This identity has the same form as (a), and the argument of (b) uses nothing else.
\end{proof}

\begin{proof}[Proof of Proposition~\ref{prop:example}]
Write $\E_{s}$ for the expectation given $\cF_{s}$. Recall that $\cF=\cF_{t}\subseteq\cG=\cG_{t+1}\subseteq\cF_{t+1}$.

We first derive expected returns. For any $s$, $\E_{s}[\eta_{s+1}]=0$, and because $\cF_{s}\subseteq\cG_{s+1}$, iterated expectations give $\E_{s}[\eta^{w}_{s+1}]=\E_{s}[\eta_{s+1}]=0$; also $\E_{s}[\varsigma_{s+1}]=0$. Since the states are $\cF_{s}$-measurable, $\E_{s}[\hat g_{s+1}]=\phi_{g}g_{s}+\phi_{\beta}\beta_{s}$ and $\E_{s}[\lambda_{s+1}]=\phi_{\lambda}\lambda_{s}$. Substituting~\eqref{eq:pd} into the return relation,
\begin{align*}
\E_{s}[r_{s+1}]&=\text{const}+\big(\rho a_{g}\phi_{g}-a_{g}+1\big)g_{s}+\big(\rho a_{g}\phi_{\beta}-a_{g}\big)\beta_{s}+\big(a_{\lambda}-\rho a_{\lambda}\phi_{\lambda}\big)\lambda_{s}\\
&=\text{const}+\lambda_{s}-(1-\rho\phi_{\beta})\,a_{g}\,\beta_{s},
\end{align*}
using $a_{g}(1-\rho\phi_{g})=1$ and $a_{\lambda}(1-\rho\phi_{\lambda})=1$.

(a) The unexpected return is $r_{t+1}-\E_{t}[r_{t+1}]=\rho\,a_{g}\big(\eta_{t+1}-(1-\theta)\eta^{w}+\varsigma_{t+1}\big)-\rho\,a_{\lambda}\xi_{t+1}+\varepsilon_{t+1}$. Since $\E_{t}[r_{t+1}]$ is $\cF$-measurable, $c^{r}$ is the conditional expectation of this expression given $\cG$, which is $\varepsilon^{w}+\rho\,\theta\,a_{g}\eta^{w}-\rho\,a_{\lambda}\xi^{w}$.

(b) Consider first the later shocks. For $i\ge2$, each shock dated $t+i$ has zero conditional expectation given $\cF_{t+i-1}\supseteq\cG$, and so does its text-conveyed part, because $\E[\eta^{w}_{t+i}\mid\cF_{t+i-1}]=\E[\eta_{t+i}\mid\cF_{t+i-1}]=0$. These shocks therefore have zero text innovation. For $i=1$, $\Nw[\varepsilon_{t+1}]=\varepsilon^{w}$, $\Nw[\eta_{t+1}]=\Nw[\eta^{w}_{t+1}]=\eta^{w}$, $\Nw[\xi_{t+1}]=\xi^{w}$, and $\Nw[\varsigma_{t+1}]=0$. Iterating the laws of motion from the $\cF$-measurable states at $t$, for $j\ge1$,
\[
\Nw[g_{t+j}]=\phi_{g}^{j-1}\eta^{w},\qquad \Nw[\lambda_{t+j}]=\phi_{\lambda}^{j-1}\xi^{w},\qquad \Nw[\beta_{t+j}]=-(1-\theta)\phi_{\beta}^{j-1}\eta^{w}.
\]
For growth, $G_{H}=\sum_{j=1}^{H+1}(g_{t+j-1}+\varepsilon_{t+j})$, and $\Nw[g_{t}]=0$, so $c^{G}_{H}=\varepsilon^{w}+\sum_{j=2}^{H+1}\phi_{g}^{j-2}\eta^{w}$, which is the stated expression. For later returns, let $j\ge1$. Because $\cG\subseteq\cF_{t+1}\subseteq\cF_{t+j}$, iterated expectations and the formula for expected returns give $\Nw[r_{t+1+j}]=\Nw[\lambda_{t+j}]-(1-\rho\phi_{\beta})a_{g}\Nw[\beta_{t+j}]=\phi_{\lambda}^{j-1}\xi^{w}+(1-\theta)(1-\rho\phi_{\beta})\,a_{g}\,\phi_{\beta}^{j-1}\eta^{w}$. Summing over $j=1,\dots,H$ gives $c^{R}_{H}$. For valuation, by~\eqref{eq:pd} and because $\pd_{t}$ is $\cF$-measurable, $c^{V}_{H}=\Nw[\pd_{t+1+H}]=a_{g}\big(\phi_{g}^{H}-(1-\theta)\phi_{\beta}^{H}\big)\eta^{w}-a_{\lambda}\phi_{\lambda}^{H}\xi^{w}$.
\end{proof}

\begin{proposition}[Shares in the example]\label{prop:exshares}
In the example of Section~\ref{sec:example}, suppose that $\varepsilon^{w}$, $\eta^{w}$, and $\xi^{w}$ are mutually uncorrelated with variances $\sigma_{\varepsilon}^{2}$, $\sigma_{\eta}^{2}$, and $\sigma_{\xi}^{2}$, and that $\Var(c^{r})>0$. Let $\pi_{\varepsilon}=\sigma_{\varepsilon}^{2}/\Var(c^{r})$, $\pi_{\eta}=\rho^{2}\theta^{2}a_{g}^{2}\sigma_{\eta}^{2}/\Var(c^{r})$, and $\pi_{\xi}=\rho^{2}a_{\lambda}^{2}\sigma_{\xi}^{2}/\Var(c^{r})$ be the shares of the variance of $c^{r}$ due to the three kinds of news. Then for every $H\ge1$,
\begin{align*}
\omega_{G}(c^{r})&=\pi_{\varepsilon}+\pi_{\eta}\,\frac{(1-\rho\phi_{g})(1-\phi_{g}^{H})}{\rho\,\theta\,(1-\phi_{g})},\\
\omega_{R}(c^{r})&=\pi_{\xi}\,\frac{(1-\rho\phi_{\lambda})(1-\phi_{\lambda}^{H})}{\rho\,(1-\phi_{\lambda})}-\pi_{\eta}\,\frac{1-\theta}{\theta}\,\frac{(1-\rho\phi_{\beta})(1-\phi_{\beta}^{H})}{\rho\,(1-\phi_{\beta})},\\
\omega_{V}(c^{r})&=\pi_{\eta}\,\frac{\phi_{g}^{H}-(1-\theta)\phi_{\beta}^{H}}{\rho\,\theta}+\pi_{\xi}\,\frac{\phi_{\lambda}^{H}}{\rho},
\end{align*}
and the payout share is $\omega_{D}(c^{r})=1-\omega_{G}(c^{r})-\omega_{R}(c^{r})-\omega_{V}(c^{r})$. The payout share also equals $-(1-\rho)$ times the sum of the valuation shares at horizons $0$ to $H$, where the formula for $\omega_{V}$ holds at horizon $0$ as well.
\end{proposition}

\begin{proof}[Proof of Proposition~\ref{prop:exshares}]
With uncorrelated components, $\Var(c^{r})=\sigma_{\varepsilon}^{2}+\rho^{2}\theta^{2}a_{g}^{2}\sigma_{\eta}^{2}+\rho^{2}a_{\lambda}^{2}\sigma_{\xi}^{2}$ and
\begin{align*}
\Cov(c^{r},c^{G}_{H})&=\sigma_{\varepsilon}^{2}+\rho\,\theta\,a_{g}\,\frac{1-\phi_{g}^{H}}{1-\phi_{g}}\,\sigma_{\eta}^{2},\\
-\Cov(c^{r},c^{R}_{H})&=\rho\,a_{\lambda}\frac{1-\phi_{\lambda}^{H}}{1-\phi_{\lambda}}\sigma_{\xi}^{2}-\rho\,\theta(1-\theta)(1-\rho\phi_{\beta})\,a_{g}^{2}\,\frac{1-\phi_{\beta}^{H}}{1-\phi_{\beta}}\sigma_{\eta}^{2},\\
\Cov(c^{r},c^{V}_{H})&=\rho\,\theta\,a_{g}^{2}\big(\phi_{g}^{H}-(1-\theta)\phi_{\beta}^{H}\big)\sigma_{\eta}^{2}+\rho\,a_{\lambda}^{2}\phi_{\lambda}^{H}\sigma_{\xi}^{2}.
\end{align*}
Dividing by $\Var(c^{r})$ and using the definitions of $\pi_{\varepsilon}$, $\pi_{\eta}$, and $\pi_{\xi}$, together with $1/a_{g}=1-\rho\phi_{g}$ and $1/a_{\lambda}=1-\rho\phi_{\lambda}$, gives the three shares. With $D_{H}$ defined as in the text, $r_{t+1}=G_{H}+D_{H}-R_{H}+V_{H}$ identically, all terms have finite second moments, and the argument of Proposition~\ref{prop:fates}(b) shows that the four shares add to one. Since $D_{H}=\sum_{s=t+1}^{t+1+H}\big(k_{0}-(1-\rho)\pd_{s}\big)$, also $c^{D}_{H}=-(1-\rho)\sum_{j=0}^{H}\Nw[\pd_{t+1+j}]$, where $\Nw[\pd_{t+1+j}]$ is given by the expression for $c^{V}_{j}$, which holds for $j=0$ as well. Taking the covariance with $c^{r}$ and dividing by $\Var(c^{r})$ gives the relation between the payout share and the valuation shares.
\end{proof}

When the reaction is complete and the text carries no discount-rate news, $\omega_{V}(c^{r})=(\pi_{\eta}/\rho)\,\phi_{g}^{H}$. If the text carries growth news and no discount-rate news ($\sigma^{2}_{\eta}>0=\sigma^{2}_{\xi}$), $\omega_{R}(c^{r})$ has the sign of $\theta-1$ at every horizon, and it is zero at every horizon if and only if $\theta=1$. The payout share is of order $1-\rho$ at a fixed horizon and grows in absolute value with the horizon.

\paragraph{Estimation} Let $W$ collect the conditioning variables, the state $S$ and a representation of the text, and let $m^{k}(W)=\E[Y^{k}\mid W]$.

\begin{proposition}[Orthogonal score]\label{prop:score}
Let $\Gamma^{kl}=\E[m^{k}(W)\,m^{l}(W)]$ and, for candidate functions $\hat m=(\hat m^{k},\hat m^{l})$, let
\[
\psi(Z;\hat m)=Y^{k}\,\hat m^{l}(W)+Y^{l}\,\hat m^{k}(W)-\hat m^{k}(W)\,\hat m^{l}(W),\qquad Z=(Y^{k},Y^{l},W).
\]
\begin{enumerate}[label=(\alph*)]
\item For any fixed square-integrable $\hat m$, $\E[\psi(Z;\hat m)]=\Gamma^{kl}-\E\big[(\hat m^{k}-m^{k})(\hat m^{l}-m^{l})\big]$.
\item Let $Z_{1},\dots,Z_{n}$ be independent and identically distributed, split into a fixed number of folds $I_{j}$ of proportional size, and let $\hat m_{(j)}$ be estimated without fold $j$. Let $\hat\Gamma^{kl}=n^{-1}\sum_{j}\sum_{i\in I_{j}}\psi(Z_{i};\hat m_{(j)})$. If $\E[(Y^{k})^{2}\mid W]$, $|m^{k}|$, and $|\hat m^{k}|$ are bounded by a constant for both indices, $\|\hat m^{k}-m^{k}\|_{2}=o_{p}(1)$ for both indices and $\|\hat m^{k}-m^{k}\|_{2}\,\|\hat m^{l}-m^{l}\|_{2}=o_{p}(n^{-1/2})$, where $\|\cdot\|_{2}$ is the $L^{2}$ norm over a new draw of $W$ with the fitted function held fixed, then $\sqrt{n}\,(\hat\Gamma^{kl}-\Gamma^{kl})$ converges in distribution to $N\big(0,\Var(\psi(Z;m))\big)$.
\item Let $m_{0}$ and $m_{1}$ be the conditional means given $S$ and given $(S,x)$, and let $c=m_{1}-m_{0}$ and $\hat c=\hat m_{1}-\hat m_{0}$. Then $\Cov(c^{k},c^{l})=\E[m_{1}^{k}m_{1}^{l}]-\E[m_{0}^{k}m_{0}^{l}]$, and the difference of the two scores equals
\[
(Y^{k}-\hat m_{0}^{k})\,\hat c^{l}+(Y^{l}-\hat m_{0}^{l})\,\hat c^{k}-\hat c^{k}\,\hat c^{l}.
\]
If $\E[\hat c^{k}\mid S]=\E[\hat c^{l}\mid S]=0$, the expectation of this difference for fixed $\hat m$ is $\Cov(c^{k},c^{l})-\E\big[(\hat c^{k}-c^{k})(\hat c^{l}-c^{l})\big]$.
\end{enumerate}
\end{proposition}

Part (a) is the reason the score is useful: the error from replacing the true conditional means by fitted ones is the product of two estimation errors. By part (c), the estimate of a concept's own variance falls short of the truth by the mean squared error of the fitted concept. Part (b) is the standard result for cross-fitted estimators \citep{chernozhukov2018double}, applied to a functional of the type studied by \citet{williamson2021nonparametric}. Its limit is degenerate when both concepts are zero.

\begin{proof}[Proof of Proposition~\ref{prop:score}]
(a) By iterated expectations $\E[Y^{k}\hat m^{l}]=\E[m^{k}\hat m^{l}]$ and $\E[Y^{l}\hat m^{k}]=\E[m^{l}\hat m^{k}]$, so $\E[\psi(Z;\hat m)]=\E[m^{k}\hat m^{l}+m^{l}\hat m^{k}-\hat m^{k}\hat m^{l}]=\E[m^{k}m^{l}]-\E[(\hat m^{k}-m^{k})(\hat m^{l}-m^{l})]$.

(b) Fix a fold $j$ with $n_{j}$ observations and condition on the data outside it, so that $\hat m=\hat m_{(j)}$ is fixed. Write
\[
\frac{1}{n_{j}}\sum_{i\in I_{j}}\big(\psi(Z_{i};\hat m)-\Gamma^{kl}\big)=\frac{1}{n_{j}}\sum_{i\in I_{j}}\big(\psi(Z_{i};m)-\Gamma^{kl}\big)+A_{j}+B_{j},
\]
where $B_{j}=\E[\psi(Z;\hat m)\mid\hat m]-\Gamma^{kl}$ and $A_{j}$ is the average over the fold of $\psi(Z_{i};\hat m)-\psi(Z_{i};m)$ minus its conditional mean. By part (a) and the Cauchy--Schwarz inequality, $|B_{j}|\le\|\hat m^{k}-m^{k}\|_{2}\|\hat m^{l}-m^{l}\|_{2}=o_{p}(n^{-1/2})$. For $A_{j}$, note that
\[
\psi(Z;\hat m)-\psi(Z;m)=Y^{k}(\hat m^{l}-m^{l})+Y^{l}(\hat m^{k}-m^{k})-\hat m^{k}(\hat m^{l}-m^{l})-m^{l}(\hat m^{k}-m^{k}),
\]
so under the boundedness conditions its conditional second moment is at most $\varrho_{n}=4(C+C^{2})\big(\|\hat m^{k}-m^{k}\|_{2}^{2}+\|\hat m^{l}-m^{l}\|_{2}^{2}\big)=o_{p}(1)$, where $C$ is the bound. Conditionally on $\hat m$, $\sqrt{n_{j}}A_{j}$ has mean zero and variance at most $\varrho_{n}$, so by Chebyshev's inequality $\Pr\big(\sqrt{n_{j}}|A_{j}|>\epsilon\mid\hat m\big)\le\min\{1,\varrho_{n}/\epsilon^{2}\}$ for every $\epsilon>0$. The right-hand side is bounded by one and converges to zero in probability, so its expectation converges to zero and $\sqrt{n_{j}}A_{j}=o_{p}(1)$. Summing over the finitely many folds, $\sqrt{n}(\hat\Gamma^{kl}-\Gamma^{kl})=n^{-1/2}\sum_{i}(\psi(Z_{i};m)-\Gamma^{kl})+o_{p}(1)$, and the central limit theorem applies because $\psi(Z;m)$ has finite variance.

(c) Since $\E[m_{1}^{k}\mid S]=m_{0}^{k}$, $\E[m_{1}^{k}m_{0}^{l}]=\E[m_{0}^{k}m_{0}^{l}]=\E[m_{0}^{k}m_{1}^{l}]$, and expanding $\E[(m_{1}^{k}-m_{0}^{k})(m_{1}^{l}-m_{0}^{l})]$ gives the first claim; concepts have mean zero, so this is their covariance. For the second, substitute $\hat m_{1}=\hat m_{0}+\hat c$ into the difference of the two scores: $Y^{k}\hat c^{l}+Y^{l}\hat c^{k}-(\hat m_{0}^{k}+\hat c^{k})(\hat m_{0}^{l}+\hat c^{l})+\hat m_{0}^{k}\hat m_{0}^{l}=(Y^{k}-\hat m_{0}^{k})\hat c^{l}+(Y^{l}-\hat m_{0}^{l})\hat c^{k}-\hat c^{k}\hat c^{l}$. For the third, fix $\hat m$. By iterated expectations given $(S,x)$, $\E[(Y^{k}-\hat m_{0}^{k})\hat c^{l}]=\E[c^{k}\hat c^{l}]+\E[(m_{0}^{k}-\hat m_{0}^{k})\hat c^{l}]$, and the last term is zero because $m_{0}^{k}-\hat m_{0}^{k}$ is a function of $S$ and $\E[\hat c^{l}\mid S]=0$. The same holds with $k$ and $l$ exchanged, so the expectation of the difference is $\E[c^{k}\hat c^{l}+c^{l}\hat c^{k}-\hat c^{k}\hat c^{l}]=\E[c^{k}c^{l}]-\E[(\hat c^{k}-c^{k})(\hat c^{l}-c^{l})]$.
\end{proof}

\paragraph{Shares as ratios} A share of the text part is a ratio of two such estimates. Suppose that the fitted concepts of two outcomes are least-squares fits on the same regressors in the same training sample, and that the parts of the outcomes that the concepts leave unexplained, $u^{k}=\tilde Y^{k}-c^{k}$, are homoskedastic and uncorrelated across observations given the regressors. The two estimation errors are then the projections of $u^{k}$ and $u^{l}$ on the regressors, and
\[
\E\big[(\hat c^{k}-c^{k})(\hat c^{l}-c^{l})\big]=\mathrm{MSE}_{k}\,\frac{\Cov(u^{k},u^{l})}{\Var(u^{k})},\qquad \mathrm{MSE}_{k}=\E\big[(\hat c^{k}-c^{k})^{2}\big].
\]
Let $k$ be the event return and write $\omega(a)=\Cov(a,\tilde Y^{l})/\Var(a)$ for the share of component $a$ in outcome $l$. Because $u^{r}$ is uncorrelated with $c^{l}$, $\Cov(u^{r},u^{l})/\Var(u^{r})=\omega(u^{r})$, the share of the remainder. By part (c), the expected numerator and denominator of the estimated share of the text part are $\Var(c^{r})\,\omega(c^{r})-\mathrm{MSE}_{r}\,\omega(u^{r})$ and $\Var(c^{r})-\mathrm{MSE}_{r}$, and their ratio is
\[
\omega(u^{r})+\big[\omega(c^{r})-\omega(u^{r})\big]\,\frac{\Var(c^{r})}{\Var(c^{r})-\mathrm{MSE}_{r}}.
\]
The estimated share is displaced away from the share of the remainder, and it is not displaced when the two shares are equal. For the ridge fits that we use the statement is an approximation.

\paragraph{Readouts fitted within a class} A readout is fitted within a class, and the class may not contain the true conditional expectation.

\begin{proposition}[Projected concepts]\label{prop:readable}
Let $\cL$ be a closed linear subspace of $L^{2}(\cG)$ that contains $L^{2}(\cF)$, let $\Pi$ be the orthogonal projection onto $\cL$, and define the \emph{projected concept} $c^{k}_{\cL}=\Pi[Y^{k}]-\E[Y^{k}\mid\cF]$.
\begin{enumerate}[label=(\alph*)]
\item $c^{k}_{\cL}=\Pi[c^{k}]$, $\E[c^{k}_{\cL}\mid\cF]=0$, and $\Var(c^{k})=\Var(c^{k}_{\cL})+\E\big[(c^{k}-c^{k}_{\cL})^{2}\big]$.
\item $\E[\tilde Y^{k}q]=\E[c^{k}_{\cL}\,q]$ for every $q\in\cL$. In particular, $\E\big[\tilde Y^{k}c^{l}_{\cL}+\tilde Y^{l}c^{k}_{\cL}-c^{k}_{\cL}c^{l}_{\cL}\big]=\Cov(c^{k}_{\cL},c^{l}_{\cL})$.
\item If $a^{\top}Y$ is $\cF$-measurable, then $a^{\top}c_{\cL}=0$. Consequently, when the same class is used for every outcome, $c^{r}_{\cL}=c^{G}_{H,\cL}+c^{D}_{H,\cL}-c^{R}_{H,\cL}+c^{V}_{H,\cL}$, and if $\Var(c^{r}_{\cL})>0$ the shares of the projected text part, $\Cov(c^{r}_{\cL},\tilde G_{H})/\Var(c^{r}_{\cL})$ and its three analogues, add to one.
\item If $c^{r}\in\cL$, the shares of $c^{r}_{\cL}$ equal those of $c^{r}$. Otherwise they can differ. In the example of Section~\ref{sec:example} with $\theta=1$, $0<\phi_{g}<1$, $\phi_{\lambda}\ge0$, uncorrelated $(\varepsilon^{w},\eta^{w},\xi^{w})$ and $\sigma^{2}_{\varepsilon},\sigma^{2}_{\eta}>0$, the class $\cL=L^{2}(\cF)\oplus\operatorname{span}\{\varepsilon^{w}\}$ gives $c^{r}_{\cL}=\varepsilon^{w}$, whose growth share is one and whose valuation share is zero, whereas $\omega_{V}(c^{r})>0$.
\end{enumerate}
\end{proposition}

The proposition qualifies Proposition~\ref{prop:concept}(iii): a projected concept is guaranteed to be unchanged by a change of coordinates only if the change leaves the class unchanged. The same statements hold, with the same proof, when conditional expectation given $\cF$ is replaced by projection on a closed subspace of $\cL$ that contains the constants, such as the span of a finite basis in the state variables that includes a constant.

\begin{proof}[Proof of Proposition~\ref{prop:readable}]
(a) For $q\in\cL\subseteq L^{2}(\cG)$, $\E[Yq]=\E[\E[Y\mid\cG]q]$, so $\Pi[Y]=\Pi[\E[Y\mid\cG]]$. Since $\E[Y\mid\cF]\in L^{2}(\cF)\subseteq\cL$, $\Pi[\E[Y\mid\cF]]=\E[Y\mid\cF]$. Hence $c^{k}_{\cL}=\Pi\big[\E[Y^{k}\mid\cG]-\E[Y^{k}\mid\cF]\big]=\Pi[c^{k}]$. For $q\in L^{2}(\cF)\subseteq\cL$, $\E[c^{k}_{\cL}q]=\E[c^{k}q]=0$, so $\E[c^{k}_{\cL}\mid\cF]=0$; in particular $c^{k}_{\cL}$ has mean zero, and the variance identity is the Pythagorean theorem. (b) For $q\in\cL$, $\E[\tilde Y^{k}q]=\E[\Pi[\tilde Y^{k}]q]$, and $\Pi[\tilde Y^{k}]=\Pi[Y^{k}]-\E[Y^{k}\mid\cF]=c^{k}_{\cL}$. Taking $q=c^{l}_{\cL}$ and then exchanging the roles of $k$ and $l$ gives $\E[\tilde Y^{k}c^{l}_{\cL}]=\E[\tilde Y^{l}c^{k}_{\cL}]=\E[c^{k}_{\cL}c^{l}_{\cL}]$. (c) $\Pi$ is linear and leaves $L^{2}(\cF)$ fixed, so $a^{\top}c_{\cL}=\Pi[a^{\top}Y]-\E[a^{\top}Y\mid\cF]=0$ when $a^{\top}Y$ is $\cF$-measurable. Applied to the identity of Proposition~\ref{prop:fates}(a), this gives the identity for projected concepts. For the shares, part (b) with $q=c^{r}_{\cL}$ gives $\E\big[c^{r}_{\cL}(\tilde G_{H}+\tilde D_{H}-\tilde R_{H}+\tilde V_{H})\big]=\E[c^{r}_{\cL}\tilde r^{E}]=\Var(c^{r}_{\cL})$. (d) If $c^{r}\in\cL$, then $c^{r}_{\cL}=c^{r}$. In the example, $\varepsilon^{w}$ is orthogonal to $L^{2}(\cF)$ because $\E[\varepsilon^{w}\mid\cF]=0$, so the projection of $c^{r}$ on $\cL$ is $\big(\E[c^{r}\varepsilon^{w}]/\sigma^{2}_{\varepsilon}\big)\varepsilon^{w}=\varepsilon^{w}$ by Proposition~\ref{prop:example}(a). As in the proof of Proposition~\ref{prop:fates}(b), $\Cov(\varepsilon^{w},\tilde G_{H})=\E[\varepsilon^{w}c^{G}_{H}]=\sigma^{2}_{\varepsilon}$ and $\Cov(\varepsilon^{w},\tilde V_{H})=\E[\varepsilon^{w}c^{V}_{H}]=0$ by Proposition~\ref{prop:example}(b), whereas $\omega_{V}(c^{r})=(\pi_{\eta}\phi_{g}^{H}+\pi_{\xi}\phi_{\lambda}^{H})/\rho>0$ by Proposition~\ref{prop:exshares}.

None of these steps uses a property of conditional expectation given $\cF$ other than that it is the orthogonal projection onto a closed subspace of $\cL$ that contains the constants. The statements therefore hold when $L^{2}(\cF)$ is replaced by any closed subspace $\cL_{0}\subseteq\cL$ that contains the constants, with $\tilde Y$ the residual from the projection on $\cL_{0}$.
\end{proof}

\paragraph{Nested representations} A representation that is a function of another carries less information, and the loss can be measured in the units of any outcome.

\begin{proposition}[Information accounting]\label{prop:info}
Let $y$ be square integrable with $\Var\big(y-\E[y\mid\cF]\big)>0$, and let $\cF\subseteq\cH_{1}\subseteq\cH_{2}$. Write $m_{j}=\E[y\mid\cH_{j}]$ and $R^{2}_{j}=\Var\big(m_{j}-\E[y\mid\cF]\big)/\Var\big(y-\E[y\mid\cF]\big)$. Then
\[
R^{2}_{2}=R^{2}_{1}+\Delta,\qquad \Delta=\frac{\E\big[(m_{2}-m_{1})^{2}\big]}{\Var\big(y-\E[y\mid\cF]\big)}\ \ge\ 0,
\]
and the following are equivalent: (a) $\Delta=0$; (b) $m_{2}=m_{1}$ almost surely; (c) $\E\big[(y-m_{1})\,q\big]=0$ for every bounded $\cH_{2}$-measurable $q$.
\end{proposition}

\begin{proof}[Proof of Proposition~\ref{prop:info}]
Let $m_{0}=\E[y\mid\cF]$. Since $\cH_{1}\subseteq\cH_{2}$, $\E[m_{2}\mid\cH_{1}]=m_{1}$, so $m_{2}-m_{1}$ is orthogonal to $L^{2}(\cH_{1})$, which contains $m_{1}-m_{0}$. Hence $\E[(m_{2}-m_{0})^{2}]=\E[(m_{1}-m_{0})^{2}]+\E[(m_{2}-m_{1})^{2}]$; both $m_{j}-m_{0}$ have mean zero, and dividing by $\Var(y-m_{0})$ gives the identity. (a) and (b) are equivalent by the definition of $\Delta$. If (b) holds, then for bounded $\cH_{2}$-measurable $q$, $\E[(y-m_{1})q]=\E[(y-m_{2})q]=0$, which is (c). Conversely, let $q_{n}=(m_{2}-m_{1})\mathbf{1}\{|m_{2}-m_{1}|\le n\}$, which is bounded and $\cH_{2}$-measurable. Then (c) gives $0=\E[(y-m_{1})q_{n}]=\E[(m_{2}-m_{1})q_{n}]=\E[(m_{2}-m_{1})^{2}\mathbf{1}\{|m_{2}-m_{1}|\le n\}]$ for every $n$, so $m_{2}=m_{1}$ almost surely.
\end{proof}

\clearpage
\section{Estimation Details}\label{app:est}

\subsection{Text}

The text of a filing is its item section, from the first item heading to the signature block, followed by the text of any press release furnished as Exhibit 99. When a filing has such exhibits we keep the first 2,500 characters of the item section, and otherwise the first 12,000. The title is the first line of the first press release that reads as a headline or, when there is no press release, the first 40 words of the item section. A firm-day is assigned the first of the following groups that applies to any of its filings: results of operations (Item 2.02); acquisitions and changes of control (2.01, 5.01); material agreements (1.01, 1.02, without 2.03 or 3.02); financing, listing, and bankruptcy (1.03, 2.03, 2.04, 3.01 to 3.03); restructuring, impairments, and auditors (2.05, 2.06, 4.01, 4.02); officers and directors (5.02); governance and votes (5.03 to 5.08); Regulation FD disclosure (7.01); and other events (8.01).

A filing is represented by the average of the language model's final-layer vectors over its first 2,048 tokens, a vector of dimension 4,096, and a firm-day by the average over its filings. Each coordinate is centered and scaled with the mean and standard deviation of January--March 2014. The novelty embedding is the firm-day vector less an exponentially weighted average of the same firm's vectors on earlier event days, with weight 0.1 on the most recent day, shrunk toward the mean of January--March 2014 when the firm has few earlier filings. The state includes the first 64 principal components of that average, with directions estimated on the filings of 2014.

The encoder is the checkpoint \texttt{Diamegs/PIT-4B-201312} in the Hugging Face repository of the model, at revision \texttt{6a3b2e1}, with the tokenizer distributed with it. We run it in evaluation mode in 16-bit floating point with version 4.48.3 of the transformers library and PyTorch 2.4, and we average the final-layer vectors over the tokens of a filing with equal weights. The SHA-256 digests of the model's two code files begin \texttt{c9f44711} and \texttt{b10a78f6}.

\subsection{Timing of the State Variables}

Table~\ref{tab:timing_state} lists the variables of the state and the date at which each is measured.

\begin{table}[tbp]
\caption{State Variables and the Information They Use}\label{tab:timing_state}
\tabwrap{\begin{tabular}{@{}p{0.33\textwidth}p{0.62\textwidth}@{}}
\toprule
Variable & Timing \\
\midrule
Market value, stock price, returns over 1, 5, 21, 63, and 252 days, volatility over 21 and 63 days, turnover, illiquidity \citep{amihud2002illiquidity}, dividend yield & CRSP, through the close of day $d-1$ \\\addlinespace[2pt]
Institutional ownership & Latest quarterly 13F holdings dated at least 50 days before day $d$ \\\addlinespace[2pt]
Numbers of recent event days, filings, and analyst revisions; days since the last earnings announcement & Through day $d-1$ \\\addlinespace[2pt]
Days to the projected next earnings announcement; indicators for a projected date within one day and for an overdue announcement & Projected date is 364 days after an announcement that was public before day $d$; no realized future date is used \\\addlinespace[2pt]
Market returns over 1, 5, and 21 days and market volatility, with squares & Through the close of day $d-1$; standardized on January--March 2014 \\\addlinespace[2pt]
Indicators for 50 two-digit industries & Industries chosen on January--March 2014 \\\addlinespace[2pt]
Principal components of the firm's past text & Filings before day $d$; directions estimated on 2014 \\\addlinespace[2pt]
Book-to-market, return on equity, asset growth, leverage & Quarters whose results were released before the close of day $d-1$; balance-sheet items dated no earlier than 45 days after the end of fiscal quarters 1--3 and 90 days after the fiscal year end \\\addlinespace[2pt]
Forecast earnings yield for the next two fiscal years, dispersion and number of forecasts & I/B/E/S forecasts time-stamped before 4 p.m.\ of day $d-1$; the fiscal years are those that follow the last fiscal year reported by then \\\addlinespace[2pt]
Return implied by price targets, number of targets, forecast of long-term growth & I/B/E/S, time-stamped before 4 p.m.\ of day $d-1$ \\\addlinespace[2pt]
Option-implied volatility & OptionMetrics, close of day $d-1$ \\
\bottomrule
\end{tabular}}
\floatnote{The table lists the state variables and the date at which each is measured. CRSP is the Center for Research in Security Prices, I/B/E/S the Institutional Brokers' Estimate System, and 13F the quarterly report of institutional holdings on Form 13F. Continuous firm-level variables enter through piecewise-linear functions of their percentile in the distribution of the same variable over the firm-days of the three preceding calendar months, with knots at the 10th, 25th, 50th, 75th, and 90th percentiles and with indicators for missing values. The last four rows enter only for the outcomes that involve fundamentals. The accounting data are from the standard Compustat quarterly file, which can contain values restated after the original release.}
\end{table}

\subsection{Fundamentals}

A fiscal quarter becomes known at the close of the first trading day whose close follows the release of its results. When the Compustat report date coincides with an I/B/E/S announcement date for which a time is recorded, the release time is the I/B/E/S time. Otherwise the quarter is dated on the trading day after the report date, and when there is no report date, 90 days after the quarter ends. Sales are summed over the trailing four quarters and must be positive. The fundamental before the event is the latest value known at the close of day $d-1$, and the fundamental at horizon $H$ is the latest value known at the close of day $d+H$. The valuation ratio is the log of market capitalization over the fundamental, which equals the log of the price over the fundamental per share.

The growth of trailing sales between two dates is the sum of the changes that occur when each new quarter becomes known. Let $S_{0}$ be trailing sales known at the close of day $d-1$ and $S_{1},\dots,S_{J}$ their values after each of the $J$ quarters that become known by the close of day $d+H$. Then
\[
\log\frac{S_{J}}{S_{0}}=\sum_{j=1}^{J}\log\frac{S_{j}}{S_{j-1}},
\]
and step $j$ is assigned to the quarter that enters at that step. When a quarter enters, the same fiscal quarter of the previous year leaves the sum, so a step equals $\log\{1+(s_{j}-s_{j}^{-})/S_{j-1}\}$, where $s_{j}$ is the sales of the quarter that enters and $s_{j}^{-}$ the sales of the quarter that leaves; the step is the log of one plus the change in quarterly sales from a year earlier relative to trailing sales. The quarter that leaves was reported about a year before, so the news in a step is the news of the quarter that enters. We group the quarters that become known between the close of day $d-1$ and the close of day $d+H$ by when they were earned: announced on the event day; ended before the event day and reported after it; in progress on the event day; and begun after the event day. With the change in the number of shares, the five pieces sum exactly to the growth of sales per share. On announcement days the quarter announced is the only one to enter trailing sales within a month for \N{auOnlyQ}\% of firm-days. On \N{auOtherAnn}\% of other event days Compustat dates a release to the event day although I/B/E/S records no announcement.

The most recent fiscal quarter in the trailing sales known before the event ended a median of \N{auAgeMed} days earlier, and no more than \N{auAgeNinety} days earlier on 90\% of firm-days. Daily gross returns are bounded below at 0.01 before logs are taken; since 2014 the bound is reached on \N{auFloor} of \N{auDaily} million daily observations and on no day in the event or outcome windows of the sample. A stock is delisted, for any reason including a merger, within a week of the event day on \N{auDelW}\% of firm-days and within a year on \N{auDelY}\%. Table~\ref{tab:stab} reports the main contributions without the firm-days that precede a delisting and without those with the most extreme returns and growth rates.

\subsection{Readout}

The state is removed from the outcomes and from the embedding by least squares on the functions described in the note to Table~\ref{tab:timing_state}, and the readout is a ridge regression of the residual outcome on the residual embedding. Two optional adjustments let the response to a given text differ across firms. With the volatility adjustment, outcomes are divided by $s=\sigma/0.02$ before the readout is fitted, where $\sigma$ is the firm's daily return volatility over the previous 63 days, bounded between 0.5\% and 10\%, and fitted values are multiplied by $s$. With group scales, the fitted values are multiplied by a factor that differs across nine groups, formed by size third and by the kind of event day: a day with a filing that reports results under Item 2.02, another day on which a filing furnishes a press release, and a day without one. The scale of each group is the slope of the unexpected outcome on the fitted concept in the months available for the quarter, bounded between 0 and 3. Table~\ref{tab:design} compares the four combinations in the chronological fit. The design with group scales and without the volatility adjustment has the highest text share in 2015--2016, \N{desNoYesEarly}\% against \N{desYesYesEarly}\% for the design with both, and we use it in both fits.

\subsection{The Two Fits}

The blocks of the cross-fit, or folds, are the 39 calendar quarters from April 2014 to December 2023. The gap is 2 months for outcomes with horizons of up to 10 trading days, 5 months up to 63 days, 14 months up to 252 days, 26 months up to 504 days, and 38 months beyond. A calendar month enters the estimation of an outcome only if the outcome is observed for every firm-day of the sample in that month, which removes the last months of the data at long horizons.

In the baseline fit, the training sample for a quarter is all other months outside the gap on either side. For each of the five gaps we fit a separate readout of the event return that excludes the same months, and every cross moment of the event return with an outcome at a given horizon uses the readout with that horizon's gap. Each outcome has one ridge penalty, chosen by out-of-sample fit on alternate quarters of 2014--2019. Group scales use all months outside the gap.

In the chronological fit, the training sample for a quarter is the months before it, less the gap of the outcome, and at least six months are required. Because no later month enters, the readout of the event return needs no gap of its own. The penalty of an outcome for a quarter is the one with the best cumulative out-of-sample score over the earlier quarters whose outcome windows have closed before the quarter starts. Until four such quarters exist, the penalty of the event return is used, and before that the largest penalty of the grid. Group scales use earlier months, less the gap, and the out-of-sample predictions that the quarter's penalty would have produced in those months. Table~\ref{tab:chronorules} lists the data that each step uses.

\begin{table}[tbp]
\caption{Data Used at Each Step of the Chronological Fit}\label{tab:chronorules}
\tabwrap{\begin{tabular}{@{}p{0.36\textwidth}p{0.59\textwidth}@{}}
\toprule
Step & Data \\
\midrule
Scaling of the embedding coordinates and of the market states; set of industries & January--March 2014 \\\addlinespace[2pt]
Novelty embedding and principal components of past text & The firm's filings on earlier event days; directions of the components from the filings of 2014 \\\addlinespace[2pt]
Percentiles of firm-level state variables & Firm-days of the three calendar months before the month of the event \\\addlinespace[2pt]
State regression and readout of an outcome with horizon $H$ & Firm-days of the months before month $q-g(H)$, at least six months \\\addlinespace[2pt]
Ridge penalty of the outcome & Out-of-sample fit in the earlier quarters that end before month $q-g(H)$; until four such quarters exist, the penalty of the event return \\\addlinespace[2pt]
Group scales of the outcome & Unexpected outcomes and out-of-sample fitted concepts, under the quarter's penalty, in the months before month $q-g(H)$ \\\addlinespace[2pt]
Design & Text share of the event return in 2015--2016 \\\addlinespace[2pt]
Fitted values for the quarter & The estimates above, fixed for the three months of the quarter \\
\bottomrule
\end{tabular}}
\floatnote{The table lists, for a calendar quarter that begins in month $q$, the data that each step of the chronological fit uses. The gap $g(H)$ of an outcome with a horizon of $H$ trading days is zero for the event return, 2 months for $H\le 10$, 5 months for $H\le 63$, 14 months for $H\le 252$, 26 months for $H\le 504$, and 38 months beyond, so that every outcome window in the estimation sample has closed before month $q$.}
\end{table}

\subsection{Standard Errors and the Bootstrap}

The orthogonal-score result of Proposition~\ref{prop:score} assumes independent observations, an assumption the data do not meet: firms recur, shocks are common to many firms, and outcome windows overlap. Four things support the inference, and each has a limit. The gap keeps the outcome windows of the estimation sample from overlapping the quarter being predicted; it does not remove the fact that the same firms appear on both sides of a fold. The standard errors described below allow for dependence within a firm and across firms on the same and nearby days; at a horizon of one year the decade contains about ten nonoverlapping periods, and at two years five, so they are less reliable there than at short horizons. The bootstrap, the jackknife, and the placebo of Tables~\ref{tab:boot}, \ref{tab:sesens}, and~\ref{tab:placebo} check the standard errors without the same asymptotic argument; the bootstrap and the jackknife hold the fitted readouts fixed. And in the simulations of Appendix~\ref{app:mc}, which cover horizons up to a month and a simpler dependence structure than the data, nominal 95\% intervals cover the true value in 90.0\% to 96.5\% of replications when the class of readouts is correct.

Let $\psi_{it}$ denote a score for firm $i$ on trading day $t$, less its sample mean, with $\psi_{it}=0$ when the firm has no event on that day, and let $n$ be the number of firm-days. With $U_{i}=\sum_{t}\psi_{it}$ and $U_{t}=\sum_{i}\psi_{it}$, the variance of the mean of the score is
\[
\hat V=\frac{1}{n^{2}}\Bigl[\sum_{i}U_{i}^{2}+\sum_{t}U_{t}^{2}+2\sum_{l=1}^{L}\Bigl(1-\frac{l}{L+1}\Bigr)\sum_{t}U_{t}U_{t-l}-\sum_{i}\sum_{t}\psi_{it}^{2}\Bigr].
\]
The first term clusters by firm, the second and third are the estimator of \citet{driscoll1998consistent} with a Bartlett kernel, and the last removes the terms that the first two count twice \citep{thompson2011simple}. The lag length $L$ is the horizon of the outcome, up to 300 trading days; at the two-year horizon this cap omits dependence across firms at longer lags.

By the delta method, a ratio of two means, $\bar N/\bar D$, has the linearized score $\{N-(\bar N/\bar D)D\}/\bar D$. For the test of proportionality, let $a$ be the score of the covariance between the text part of the event return and the concept of a later outcome, $b$ the score of the variance of the text part, and $c$ and $d$ the products of the unexpected event return with the unexpected later outcome and with itself. The contribution of the text part is $K=\bar a/\bar d$, the text share is $\tau=\bar b/\bar d$, and the share of the whole move is $W=\bar c/\bar d$. The test statistic is based on $\Delta=K-\tau W$, the difference between the contribution and the one that the text part would make in proportion to its variance, whose linearized score is
\[
\frac{1}{\bar d}\bigl[(a-Kd)-\tau(c-Wd)-W(b-\tau d)\bigr].
\]
A $z$-statistic in the paper is $\Delta$ divided by the square root of $\hat V$ for this score. Later returns enter $a$ and $c$ with the sign reversed.

The bootstrap resamples calendar quarters in circular blocks and recomputes each statistic from the resampled quarters, with the fitted readouts held fixed. Blocks are one quarter long for horizons up to a week, two quarters for horizons of a month or a quarter, and one quarter more than the horizon beyond that, so a block is longer than the outcome window. The bootstrap therefore captures dependence across firms within a block and sampling variation over time, but not the uncertainty of the fitted readouts. Table~\ref{tab:sesens} compares the test statistics of the main contributions under other estimators of the variance: with the same firm's products at nonzero lags removed from the Driscoll--Kraay term, with fewer and with more lags, with a jackknife that deletes one calendar quarter at a time, and with the bootstrap. No alternative lowers a statistic by more than 0.10 in absolute value; with fewer lags and with the jackknife several are larger.

\clearpage
\section{Additional Tables}\label{app:tables}

This appendix collects the tables to which the text refers: the choice of design (Table~\ref{tab:design}), the adding up of the estimated shares (Table~\ref{tab:addup}), the text share by item and under other texts, embeddings, and windows (Tables~\ref{tab:types} and~\ref{tab:texts}), the levels of the shares (Table~\ref{tab:levels}), subsamples (Table~\ref{tab:week}), long horizons (Table~\ref{tab:long}), other fundamentals (Table~\ref{tab:fund}), and checks on the inference, on extreme returns, and on the stability of the estimates (Tables~\ref{tab:boot} to~\ref{tab:stab}).

\begin{table}[H]
\caption{Choice of Design on Early Data}\label{tab:design}
\tabwrap{\begin{tabular}{@{}llrrrr@{}}
\toprule
Volatility & Group & \multicolumn{2}{c}{2015--2016} & \multicolumn{2}{c}{2017--2023} \\
\cmidrule(lr){3-4}\cmidrule(l){5-6}
Adjustment & Scales & Text Share (\%) & SE & Text Share (\%) & SE \\
\midrule
Yes & Yes & 3.09 & 0.28 & 3.46 & 0.24 \\
Yes & No & 2.55 & 0.32 & 2.65 & 0.28 \\
No & Yes & 3.35 & 0.25 & 4.71 & 0.17 \\
No & No & 2.37 & 0.17 & 3.66 & 0.12 \\
\bottomrule
\end{tabular}}
\floatnote{The table reports the text share of the market-adjusted event-day return in the chronological fit, for the four designs. With the volatility adjustment, outcomes are divided by the firm's volatility over the previous 63 days before the readout is fitted, and fitted values are multiplied back. With group scales, the size of the response differs across nine groups of firm-days. The design used in the paper is the one with the highest text share in 2015--2016. SE is the standard error.}
\end{table}

\begin{table}[H]
\caption{Adding Up of the Estimated Shares}\label{tab:addup}
\tabwrap{\begin{tabular}{@{}lrrrr@{}}
\toprule
 & Text & Sum of & & Whole \\
Horizon & Share (\%) & Contributions & Difference & Move \\
\midrule
\multicolumn{5}{@{}l}{\textit{Baseline, 2014--2023}} \\
1 month & 5.46 & 5.48 & 0.02 & 1.0000 \\
1 quarter & 5.46 & 5.41 & $-$0.05 & 1.0000 \\
1 year & 5.17 & 5.18 & 0.01 & 1.0000 \\
2 years & 4.75 & 4.59 & $-$0.16 & 1.0000 \\
\addlinespace[3pt]\multicolumn{5}{@{}l}{\textit{Chronological, 2017--2023}} \\
1 month & 4.92 & 4.65 & $-$0.27 & 1.0000 \\
1 quarter & 4.92 & 4.54 & $-$0.38 & 1.0000 \\
1 year & 4.92 & 4.81 & $-$0.11 & 0.9999 \\
2 years & 4.92 & 4.51 & $-$0.41 & 0.9999 \\
\bottomrule
\end{tabular}}
\floatnote{The table compares the text share of the event return with the sum of the four contributions of the text part (through the growth of sales per share, the payout, later returns with the sign reversed, and the change in valuation), in percent of the variance of the whole unexpected event-day return. In population the two are equal. ``Whole Move'' is the sum of the four shares of the whole move, which is one in population. The sample is the firm-days with sales, and the text share is that of the readout of the event return used at the horizon.}
\end{table}

\begin{table}[H]
\caption{Text Share of Event-Return Variance by Item of the Filing}\label{tab:types}
\tabwrap{\begin{tabular}{@{}lrrrrr@{}}
\toprule
 & Firm-Days & \multicolumn{2}{c}{Baseline, 2014--2023} & \multicolumn{2}{c}{Chronological, 2017--2023} \\
\cmidrule(lr){3-4}\cmidrule(l){5-6}
Item Group & (\%) & Text Share (\%) & SE & Text Share (\%) & SE \\
\midrule
Results of operations & 32.3 & 6.10 & 0.23 & 5.35 & 0.25 \\
Material agreements & 7.1 & 11.68 & 0.63 & 9.63 & 0.60 \\
Acquisitions, changes of control & 1.7 & $-$1.06 & 1.16 & $-$0.05 & 1.14 \\
Financing, listing, bankruptcy & 6.5 & 3.50 & 0.59 & 2.93 & 0.55 \\
Restructuring, impairments, auditors & 1.1 & 5.68 & 1.32 & 4.26 & 1.14 \\
Officers and directors & 16.8 & $-$1.18 & 0.50 & 0.29 & 0.40 \\
Governance and votes & 8.0 & $-$1.21 & 0.34 & $-$0.52 & 0.26 \\
Regulation FD disclosure & 11.6 & 2.44 & 0.60 & 2.75 & 0.51 \\
Other events & 13.9 & 6.00 & 0.48 & 4.67 & 0.35 \\
\bottomrule
\end{tabular}}
\floatnote{The table reports the text share of the market-adjusted event return by item group. Firm-days are classified by the items of their filings; Appendix~\ref{app:est} lists the item numbers of each group. A negative text share means that the fitted readout does worse than no readout in that group. Firm-days whose filings list none of these items (\N{noItemShare}\% of firm-days) are not shown. SE is the standard error.}
\end{table}

\begin{table}[H]
\caption{Text Share under Other Texts, Embeddings, and Windows}\label{tab:texts}
\tabwrap{\begin{tabular}{@{}lrr@{}}
\toprule
Variant & Text Share (\%) & SE \\
\midrule
\multicolumn{3}{@{}l}{\textit{Panel A. Amount of text (market-adjusted event-day return)}} \\
Title & 1.70 & 0.11 \\
First 256 tokens & 2.45 & 0.15 \\
First 2,048 tokens (baseline) & 5.49 & 0.18 \\
\addlinespace[3pt]\multicolumn{3}{@{}l}{\textit{Panel B. Embedding (market-adjusted event-day return)}} \\
Without the novelty adjustment & 4.63 & 0.16 \\
With 1,024 nonlinear features & 5.62 & 0.18 \\
\addlinespace[3pt]\multicolumn{3}{@{}l}{\textit{Panel C. Window keyed to the acceptance date (raw returns)}} \\
Acceptance date & 4.25 & 0.20 \\
\quad Previous close to open & 6.41 & 0.30 \\
\quad Open to close & 0.19 & 0.03 \\
Next trading day & 1.78 & 0.10 \\
\quad Previous close to open & 2.29 & 0.11 \\
\quad Open to close & 0.18 & 0.03 \\
\bottomrule
\end{tabular}}
\floatnote{The table reports the text share in the baseline fit, 2014--2023, for variants of the text, the embedding, and the window. Panel A reads less of each filing. Panel B changes the embedding: without the subtraction of the firm's past text, and with 1,024 random nonlinear features appended. Panel C keys the window to the calendar date of acceptance, as a data set without acceptance times must, and reports the text share of the raw return on that date and on the next trading day. SE is the standard error.}
\end{table}

\begin{table}[H]
\caption{Shares of Price Moves: Text Part, Remainder, Whole Move, and Days without News}\label{tab:levels}
\tabwrap{\setlength{\tabcolsep}{4pt}\begin{tabular}{@{}lrrrrrrrrr@{}}
\toprule
 & \multicolumn{7}{c}{Event Days} & \multicolumn{2}{c}{No News} \\
\cmidrule(lr){2-8}\cmidrule(l){9-10}
Horizon & Text Part & SE & Remainder & SE & $z$ & Whole Move & SE & Share & SE \\
\midrule
\multicolumn{10}{@{}l}{\textit{Panel A. Offset share (negative: later returns extend the move)}} \\
1 day & $-$0.059 & 0.007 & $-$0.004 & 0.003 & $-$8.08 & $-$0.007 & 0.003 & 0.046 & 0.008 \\
1 week & $-$0.114 & 0.016 & $-$0.002 & 0.007 & $-$6.96 & $-$0.008 & 0.007 & 0.073 & 0.014 \\
1 month & $-$0.151 & 0.039 & $-$0.024 & 0.011 & $-$3.33 & $-$0.030 & 0.011 & 0.075 & 0.020 \\
1 quarter & $-$0.308 & 0.074 & $-$0.058 & 0.018 & $-$3.46 & $-$0.070 & 0.019 & 0.083 & 0.037 \\
\addlinespace[3pt]\multicolumn{10}{@{}l}{\textit{Panel B. Growth share, sales per share}} \\
1 month & 0.338 & 0.043 & 0.023 & 0.008 & 6.94 & 0.040 & 0.007 & 0.023 & 0.015 \\
1 quarter & 0.486 & 0.082 & 0.051 & 0.012 & 5.04 & 0.075 & 0.012 & 0.047 & 0.025 \\
1 year & 1.462 & 0.242 & 0.200 & 0.023 & 5.17 & 0.265 & 0.023 & 0.217 & 0.051 \\
2 years & 1.386 & 0.278 & 0.393 & 0.045 & 3.45 & 0.440 & 0.041 & 0.385 & 0.089 \\
\addlinespace[3pt]\multicolumn{10}{@{}l}{\textit{Panel C. Valuation share, price to sales}} \\
1 month & 0.826 & 0.055 & 1.005 & 0.012 & $-$3.25 & 0.995 & 0.012 & 0.888 & 0.024 \\
1 quarter & 0.850 & 0.113 & 1.008 & 0.020 & $-$1.39 & 0.999 & 0.020 & 0.851 & 0.044 \\
1 year & $-$0.260 & 0.624 & 1.033 & 0.047 & $-$2.13 & 0.966 & 0.062 & 0.802 & 0.096 \\
2 years & $-$0.787 & 0.827 & 1.005 & 0.067 & $-$2.25 & 0.920 & 0.086 & 0.737 & 0.182 \\
\bottomrule
\end{tabular}}
\floatnote{The table reports the levels of the shares of Proposition~\ref{prop:fates}: the covariance of a component of the unexpected event-day return with the unexpected part of a later outcome, divided by the variance of the component, with the sign reversed for later returns. Shares are slopes and can be negative or exceed one. The estimated share of the text part divides by the estimated variance of the text part and is displaced away from the share of the remainder when the readout is imprecise, most of all at long horizons (Section~\ref{sec:est}). The $z$-statistic is that of the difference between the text part and the remainder. The payout share, not shown, is at most \N{payMax} in absolute value. Panel A uses all firm-days and Panels B and C those with sales. The columns headed ``No News'' are for firm-days on which the firm neither filed a current report nor appeared in the news collection of \citet{dong2024fnspid} on that day or the adjacent days, having filed during the preceding quarter. The estimates are from the baseline fit, 2014--2023. SE is the standard error.}
\end{table}

\begin{table}[H]
\caption{The Week after the Event Day, by Subsample}\label{tab:week}
\tabwrap{\setlength{\tabcolsep}{3.6pt}\begin{tabular}{@{}lrrrrrrrr@{}}
\toprule
 & \multicolumn{4}{c}{Baseline, 2014--2023} & \multicolumn{4}{c}{Chronological, 2017--2023} \\
\cmidrule(lr){2-5}\cmidrule(l){6-9}
Subsample & Text Part & SE & Remainder & $z$ & Text Part & SE & Remainder & $z$ \\
\midrule
All firm-days & $-$0.114 & 0.016 & $-$0.002 & $-$6.96 & $-$0.119 & 0.021 & 0.001 & $-$5.68 \\
Smallest third & $-$0.146 & 0.019 & $-$0.003 & $-$7.27 & $-$0.154 & 0.024 & $-$0.003 & $-$6.04 \\
Middle third & $-$0.077 & 0.020 & 0.002 & $-$3.69 & $-$0.081 & 0.026 & 0.008 & $-$3.38 \\
Largest third & $-$0.047 & 0.024 & $-$0.004 & $-$1.79 & $-$0.038 & 0.032 & $-$0.002 & $-$1.15 \\
Announcement days & $-$0.143 & 0.029 & $-$0.008 & $-$4.65 & $-$0.172 & 0.034 & $-$0.003 & $-$4.92 \\
Other event days & $-$0.093 & 0.016 & 0.003 & $-$5.45 & $-$0.084 & 0.023 & 0.004 & $-$3.64 \\
Accepted before the open & $-$0.080 & 0.013 & $-$0.026 & $-$3.66 & $-$0.081 & 0.016 & $-$0.024 & $-$3.26 \\
Accepted during trading hours & $-$0.366 & 0.139 & 0.033 & $-$2.90 & $-$0.242 & 0.107 & 0.038 & $-$2.59 \\
Accepted after the close & $-$0.168 & 0.031 & 0.015 & $-$5.77 & $-$0.186 & 0.039 & 0.017 & $-$5.06 \\
No later filing within five days & $-$0.129 & 0.017 & $-$0.001 & $-$7.32 & $-$0.146 & 0.022 & 0.002 & $-$6.69 \\
A later filing within five days & $-$0.070 & 0.029 & $-$0.003 & $-$2.30 & $-$0.048 & 0.034 & $-$0.002 & $-$1.32 \\
Event days in 2014--2016 & $-$0.105 & 0.029 & $-$0.016 & $-$3.01 & -- & -- & -- & -- \\
Event days in 2017--2019 & $-$0.110 & 0.017 & $-$0.014 & $-$5.13 & $-$0.094 & 0.017 & $-$0.016 & $-$4.17 \\
Event days in 2020--2023 & $-$0.121 & 0.028 & 0.010 & $-$4.53 & $-$0.135 & 0.032 & 0.009 & $-$4.46 \\
\bottomrule
\end{tabular}}
\floatnote{The table reports the offset share five trading days after the event day, for the text part and the remainder of the market-adjusted event-day move. A negative entry means that later returns extend the move. The $z$-statistic is that of the difference between the text part and the remainder. A later filing means a current report or an earnings announcement by the firm in the five trading days after the event day; \N{nofollowShare}\% of firm-days have none. A dash marks a subsample outside the period of the chronological fit. SE is the standard error of the share of the text part.}
\end{table}

\begin{table}[H]
\caption{Long Horizons and the Gap: Offset Share of the Text Part}\label{tab:long}
\tabwrap{\begin{tabular}{@{}lrrrrrrrr@{}}
\toprule
 & \multicolumn{2}{c}{Without the Gap} & \multicolumn{2}{c}{With the Gap} & \multicolumn{2}{c}{Chronological} & \multicolumn{2}{c}{Remainder} \\
\cmidrule(lr){2-3}\cmidrule(lr){4-5}\cmidrule(lr){6-7}\cmidrule(l){8-9}
Horizon & Share & SE & Share & SE & Share & SE & Share & SE \\
\midrule
1 quarter & $-$0.36 & 0.07 & $-$0.31 & 0.07 & $-$0.26 & 0.12 & $-$0.06 & 0.02 \\
1 year & $-$0.55 & 0.24 & $-$0.06 & 0.45 & $-$0.14 & 0.57 & $-$0.24 & 0.04 \\
2 years & $-$0.58 & 0.37 & 0.55 & 0.87 & $-$0.06 & 0.66 & $-$0.42 & 0.07 \\
3 years & $-$0.64 & 0.36 & 0.27 & 0.61 & 0.50 & 0.96 & $-$0.41 & 0.12 \\
\bottomrule
\end{tabular}}
\floatnote{The table reports the offset share of the text part of the market-adjusted event-day move, for the readout of the event return fitted without the gap, with the gap (the baseline), and on earlier months only (the chronological fit), and the offset share of the remainder in the baseline fit. A negative entry means that later returns extend the move. The chronological column uses event days in 2017--2023 and the other columns event days in 2014--2023; the three-year horizon ends with the event days of 2022. SE is the standard error.}
\end{table}

\begin{table}[H]
\caption{Other Fundamentals}\label{tab:fund}
\tabwrap{\begin{tabular}{@{}lrrrrrr@{}}
\toprule
 & Text & Whole & Text & & & \\
Horizon & Share (\%) & Move & Part & SE & Fraction & $z$ \\
\midrule
\multicolumn{7}{@{}l}{\textit{Panel A. Growth of book equity per share, all event days}} \\
1 month & 5.3 & 6.30 & 1.97 & 0.24 & 0.31 & 6.75 \\
1 quarter & 5.3 & 13.22 & 1.70 & 0.26 & 0.13 & 3.58 \\
1 year & 4.7 & 52.82 & 2.78 & 0.92 & 0.05 & 0.35 \\
2 years & 4.5 & 73.40 & 4.93 & 1.97 & 0.07 & 0.89 \\
\addlinespace[3pt]\multicolumn{7}{@{}l}{\textit{Panel B. Growth of book equity per share, announcement days}} \\
1 month & 5.3 & 11.43 & 3.61 & 0.42 & 0.32 & 7.02 \\
1 quarter & 5.3 & 15.09 & 3.44 & 0.52 & 0.23 & 4.85 \\
1 year & 4.8 & 52.20 & 5.86 & 1.86 & 0.11 & 2.01 \\
2 years & 4.7 & 77.91 & 10.73 & 3.42 & 0.14 & 2.22 \\
\addlinespace[3pt]\multicolumn{7}{@{}l}{\textit{Panel C. Change in operating income over market value, all event days}} \\
1 year & 5.2 & 6.91 & 2.28 & 0.29 & 0.33 & 6.61 \\
2 years & 4.8 & 4.46 & 1.77 & 0.42 & 0.40 & 3.77 \\
\addlinespace[3pt]\multicolumn{7}{@{}l}{\textit{Panel D. Change in operating income over market value, other event days}} \\
1 year & 5.0 & 2.50 & 0.67 & 0.19 & 0.27 & 3.03 \\
2 years & 4.6 & 0.68 & 0.52 & 0.35 & 0.76 & 1.42 \\
\bottomrule
\end{tabular}}
\floatnote{The table repeats the contributions of Table~\ref{tab:contrib} for other fundamentals, in the baseline fit, 2014--2023. Panels A and B use book equity per share as the fundamental, for the firm-days with positive book equity. Panels C and D replace the growth of the fundamental by the change in trailing four-quarter operating income divided by the market value of equity before the event, bounded at 0.5 in absolute value; this outcome is not a term of the identity. SE is the standard error of the text part's contribution, and $z$ tests that the text part contributes in proportion to its variance.}
\end{table}

\begin{table}[H]
\caption{Analytic and Bootstrap Standard Errors}\label{tab:boot}
\tabwrap{\setlength{\tabcolsep}{3pt}\begin{tabular}{@{}p{0.375\textwidth}rrrrrrr@{}}
\toprule
 & Text & \multicolumn{2}{c}{SE of the Text Part} & & \multicolumn{2}{c}{$z$} & Block \\
\cmidrule(lr){3-4}\cmidrule(lr){6-7}
Outcome & Part & Analytic & Bootstrap & Excess & Analytic & Bootstrap & (Quarters) \\
\midrule
\multicolumn{8}{@{}l}{\textit{Baseline, 2014--2023}} \\
Later returns, 1 day & $-$0.31 & 0.03 & 0.03 & $-$0.28 & $-$8.29 & $-$7.35 & 1 \\
Later returns, 1 week & $-$0.61 & 0.08 & 0.08 & $-$0.57 & $-$7.13 & $-$7.13 & 1 \\
Later returns, 1 month & $-$0.78 & 0.21 & 0.25 & $-$0.62 & $-$3.23 & $-$2.98 & 2 \\
Sales growth, 1 year, all event days & 7.56 & 0.73 & 0.67 & 6.19 & 7.99 & 8.53 & 5 \\
Sales growth, 1 month, announcement days & 4.04 & 0.36 & 0.45 & 3.66 & 9.79 & 7.63 & 2 \\
Sales growth, 1 year, other event days & 4.76 & 0.94 & 0.88 & 3.53 & 3.67 & 3.96 & 5 \\
\quad Quarters begun later, 1 year & 3.58 & 0.67 & 0.63 & 3.06 & 4.73 & 4.91 & 5 \\
\quad Quarters begun later, 2 years & 4.47 & 1.05 & 0.90 & 3.59 & 3.47 & 4.34 & 9 \\
\addlinespace[3pt]\multicolumn{8}{@{}l}{\textit{Chronological, 2017--2023}} \\
Later returns, 1 day & $-$0.26 & 0.04 & 0.03 & $-$0.22 & $-$5.90 & $-$5.51 & 1 \\
Later returns, 1 week & $-$0.56 & 0.10 & 0.07 & $-$0.54 & $-$5.84 & $-$6.39 & 1 \\
Later returns, 1 month & $-$0.77 & 0.25 & 0.30 & $-$0.64 & $-$2.73 & $-$2.51 & 2 \\
Sales growth, 1 year, all event days & 7.94 & 1.03 & 0.92 & 6.55 & 5.94 & 6.43 & 5 \\
Sales growth, 1 month, announcement days & 3.72 & 0.34 & 0.39 & 3.40 & 9.72 & 8.03 & 2 \\
Sales growth, 1 year, other event days & 3.91 & 0.97 & 0.78 & 2.66 & 2.66 & 3.12 & 5 \\
\quad Quarters begun later, 1 year & 3.67 & 0.59 & 0.50 & 3.05 & 5.20 & 5.99 & 5 \\
\quad Quarters begun later, 2 years & 4.13 & 0.77 & 0.49 & 2.92 & 3.90 & 5.88 & 9 \\
\bottomrule
\end{tabular}}
\floatnote{The table compares analytic and bootstrap inference. ``Text Part'' is the contribution of the text part in percent of the variance of the whole unexpected event-day return, as in Table~\ref{tab:contrib}, and ``Excess'' is its difference from the contribution it would make in proportion to its variance. The two standard errors (SE) shown are those of the text part's contribution. The $z$-statistics divide the excess by its own analytic and bootstrap standard errors, which are not shown. ``Quarters begun later'' are fiscal quarters that began after the event day; the two indented rows are for other event days. The bootstrap resamples calendar quarters in circular blocks of the length shown, with the fitted readouts held fixed, and uses 999 draws.}
\end{table}

\begin{table}[H]
\caption{The Main Test Statistics under Other Estimators of the Variance}\label{tab:sesens}
\tabwrap{\begin{tabular}{@{}lrrrrrr@{}}
\toprule
 & & Same-Firm & Fewer & More & & \\
Outcome & Paper & Lags Removed & Lags & Lags & Jackknife & Bootstrap \\
\midrule
\multicolumn{7}{@{}l}{\textit{Baseline, 2014--2023}} \\
Later returns, 1 week & $-$7.13 & $-$7.15 & $-$8.63 & $-$7.10 & $-$7.08 & $-$7.13 \\
Sales growth, 1 year & 7.99 & 8.33 & 8.25 & 8.09 & 8.94 & 8.53 \\
\quad Quarters begun later & 6.87 & 7.14 & 7.62 & 7.26 & 9.00 & 7.26 \\
\addlinespace[3pt]\multicolumn{7}{@{}l}{\textit{Chronological, 2017--2023}} \\
Later returns, 1 week & $-$5.84 & $-$5.86 & $-$7.46 & $-$6.46 & $-$6.16 & $-$6.39 \\
Sales growth, 1 year & 5.94 & 6.05 & 6.01 & 5.84 & 6.76 & 6.43 \\
\quad Quarters begun later & 6.25 & 6.42 & 6.28 & 6.28 & 7.36 & 6.89 \\
\bottomrule
\end{tabular}}
\floatnote{The table reports the $z$-statistic of the test that the text part contributes in proportion to its variance, for the two main contributions of Table~\ref{tab:contrib} and for the contribution to one-year sales growth through fiscal quarters that began after the event day (all event days), under six estimators of the variance. ``Paper'' is the estimator of Appendix~\ref{app:est}, with 5 lags for the one-week outcome and 252 for the one-year outcomes. ``Same-Firm Lags Removed'' subtracts from it the products of the same firm's scores at nonzero lags, which the firm clusters already contain. ``Fewer Lags'' uses no lags for the one-week outcome and 126 for the one-year outcomes, and ``More Lags'' uses 21 and 504. ``Jackknife'' deletes one calendar quarter at a time. ``Bootstrap'' is the block bootstrap of Table~\ref{tab:boot}.}
\end{table}

\begin{table}[H]
\caption{Placebo: Embeddings Reassigned at Random across Firms within the Day}\label{tab:placebo}
\tabwrap{\begin{tabular}{@{}lrrrrr@{}}
\toprule
Outcome & Text Share (\%) & Whole Move & Text Part & SE & $z$ \\
\midrule
Later returns, 1 day & $-$0.001 & $-$0.71 & $-$0.0004 & 0.0004 & $-$0.81 \\
Later returns, 1 week & $-$0.001 & $-$0.77 & $-$0.0023 & 0.0010 & $-$2.24 \\
Sales growth, 1 month & $-$0.002 & 3.99 & $-$0.0060 & 0.0055 & $-$1.08 \\
Sales growth, 1 year & $-$0.001 & 26.53 & $-$0.0031 & 0.0140 & $-$0.22 \\
Sales growth, 1 month, announcement days & $-$0.004 & 6.72 & $-$0.0022 & 0.0078 & $-$0.28 \\
Sales growth, 1 year, other event days & 0.001 & 24.50 & 0.0039 & 0.0218 & 0.18 \\
\bottomrule
\end{tabular}}
\floatnote{The table repeats the baseline fit after each firm-day's embedding has been replaced by that of another firm-day drawn at random from the same trading day. Entries are as in Table~\ref{tab:contrib}; ``Text Share'' is that of the placebo readout used at the horizon, and $z$ is the contribution of the placebo text part divided by its standard error (SE). Without a gap, the text share of the placebo readout of the event-day return is \N{tsPerm}\% (standard error \N{tsPermSE}).}
\end{table}

\begin{table}[H]
\caption{Contributions with Later Outcomes Winsorized}\label{tab:winsor}
\tabwrap{\begin{tabular}{@{}lrrrr@{}}
\toprule
 & \multicolumn{2}{c}{As Observed} & \multicolumn{2}{c}{Winsorized} \\
\cmidrule(lr){2-3}\cmidrule(l){4-5}
Outcome & Whole Move & Text Part & Whole Move & Text Part \\
\midrule
\multicolumn{5}{@{}l}{\textit{Panel A. Baseline fit, 2014--2023}} \\
Later returns, 1 day & $-$0.71 & $-$0.31 & $-$0.65 & $-$0.28 \\
Later returns, 1 week & $-$0.74 & $-$0.60 & $-$0.88 & $-$0.58 \\
Later returns, 1 month & $-$2.87 & $-$0.78 & $-$3.05 & $-$0.76 \\
Later returns, 1 quarter & $-$6.60 & $-$1.59 & $-$5.62 & $-$1.49 \\
Sales growth, 1 month & 3.99 & 1.94 & 2.94 & 2.07 \\
Sales growth, 1 year & 26.31 & 7.58 & 21.96 & 7.13 \\
Sales growth, 2 years & 43.20 & 6.95 & 35.83 & 6.71 \\
Sales growth, 1 year, quarters begun after the event day & 11.74 & 4.10 & 10.40 & 3.62 \\
\addlinespace[3pt]\multicolumn{5}{@{}l}{\textit{Panel B. Chronological fit}} \\
Later returns, 1 week & $-$0.87 & $-$0.42 & $-$1.00 & $-$0.42 \\
Later returns, 1 month & $-$2.91 & $-$0.57 & $-$3.15 & $-$0.58 \\
Sales growth, 1 year & 27.34 & 7.03 & 22.99 & 6.81 \\
Sales growth, 1 year, quarters begun after the event day & 12.34 & 3.84 & 10.91 & 3.58 \\
\bottomrule
\end{tabular}}
\floatnote{The table reports the covariance of the whole unexpected event-day return, and of its text part, with later outcomes, in percent of the variance of the whole move, before and after each later outcome is winsorized at its 1st and 99th percentiles. The estimator is the simple covariance of the cross-fitted text part with the later outcome less its fitted expectation given the state, without the correction term of the orthogonal score, and the unexpected event return is that of the state regression fitted without a gap, so the entries of Panel A before winsorizing differ slightly from those of Table~\ref{tab:contrib}. Later returns enter with the sign reversed. Panel B uses all event days for which a chronological fit exists (\N{nChronoWk} firm-days at one week), a longer period than the 2017--2023 of Table~\ref{tab:contrib}, so its entries are not comparable with that table.}
\end{table}

\begin{table}[H]
\caption{Stability of the Main Contributions}\label{tab:stab}
\tabwrap{\setlength{\tabcolsep}{2pt}\begin{tabular}{@{}lrcccccc@{}}
\toprule
 & & One Quarter & One Firm & & One Month & Without & Without \\
Outcome & Estimate & Left Out & Left Out & Years & per Year & Delisted & Extremes \\
\midrule
\multicolumn{8}{@{}l}{\textit{Baseline, 2014--2023}} \\
Later returns, 1 week & $-$0.61 & $-$0.64 to $-$0.58 & $-$0.62 to $-$0.60 & 9 of 10 & -- & $-$0.60 ($-$7.1) & $-$0.81 ($-$7.1) \\
Sales growth, 1 year & 7.56 & 7.23 to 7.77 & 7.50 to 7.64 & 10 of 10 & 4.18 to 12.50 & 8.87 (8.1) & 9.10 (7.5) \\
\quad Quarters begun later & 4.03 & 3.89 to 4.16 & 3.99 to 4.09 & 10 of 10 & 2.74 to 5.73 & 4.42 (6.5) & 4.36 (5.2) \\
\addlinespace[3pt]\multicolumn{8}{@{}l}{\textit{Chronological, 2017--2023}} \\
Later returns, 1 week & $-$0.56 & $-$0.60 to $-$0.53 & $-$0.57 to $-$0.56 & 7 of 7 & -- & $-$0.56 ($-$5.8) & $-$0.78 ($-$5.8) \\
Sales growth, 1 year & 7.94 & 7.32 to 8.19 & 7.87 to 8.02 & 7 of 7 & 4.55 to 13.81 & 8.93 (5.9) & 10.19 (6.6) \\
\quad Quarters begun later & 4.50 & 4.21 to 4.65 & 4.44 to 4.57 & 7 of 7 & 2.93 to 6.73 & 4.80 (5.8) & 5.36 (6.1) \\
\bottomrule
\end{tabular}}
\floatnote{The table reports how the main contributions of the text part, in percent of the variance of the whole unexpected event-day return, vary across parts of the sample. ``Estimate'' is the contribution in the whole sample; the last row of each panel is the contribution to one-year sales growth through fiscal quarters that began after the event day, for all event days. ``One Quarter Left Out'' and ``One Firm Left Out'' give the range of the estimate when one calendar quarter or one firm is removed. ``Years'' counts the calendar years in which the difference from a proportional contribution has the sign that it has in the whole sample. ``One Month per Year'' gives the range across twelve subsamples, each made of the event days of one calendar month in every year, whose one-year outcome windows overlap by no more than about a month from one year to the next; in each of them the contribution exceeds the proportional one. The last two columns remove the firm-days whose stock is delisted within the horizon, and the firm-days on which the event return, the outcome, or another term of the identity at the horizon lies outside its 0.1st to 99.9th percentile range; the same firm-days are removed for every outcome, and the $z$-statistic of the test of proportionality is in parentheses.}
\end{table}

\clearpage
\section{Monte Carlo Evidence}\label{app:mc}

We simulate panels from the example of Section~\ref{sec:example} and apply the estimator of Section~\ref{sec:estimator} to them. A panel has 300 firms and 2,520 days, and a firm has news on a quarter of its days. The parameters are $\rho=0.96$, $\phi_{g}=0.9$, $\phi_{\lambda}=0.8$, $\phi_{\beta}=0.6$, and $\theta=0.6$. On a news day the text conveys a fixed fraction of the variance of each shock, 40\% in the baseline, which makes the text share of the event return 34\%. One-tenth of the conveyed variation and three-tenths of the rest are common to all firms on a day, and shocks have Student-$t$ distributions with five degrees of freedom. The simulated embedding has 32 coordinates. It is a random rotation of four blocks: the conveyed shocks, five coordinates of noise, four coordinates that load on the state before the event, and further noise. Outcomes are the event return and the returns over the next 1, 5, and 21 periods. Each design is replicated 200 times.

Table~\ref{tab:mc} reports the results for the text share of the event return and for the offset shares at the three horizons. The second design raises the dimension of the embedding to 256 and doubles the weight of common shocks. The next two lower the fraction of variance that the text conveys until the text share is 1.6\% and then 0.4\%. In the last design the growth shock enters the embedding only through its sign and its absolute value, as separate coordinates, so that a linear readout captures 55\% of its variance. The estimates are then centered on the shares of the projected concept of Proposition~\ref{prop:readable} and not on those of the full concept.

\begin{table}[H]
\caption{Monte Carlo Evidence on the Estimator}\label{tab:mc}
\tabwrap{\begin{tabular}{lcccc}
\toprule
 & \multicolumn{2}{c}{Text Share} & \multicolumn{2}{c}{Offset Shares} \\
\cmidrule(lr){2-3}\cmidrule(lr){4-5}
Design & Bias (\%) & Coverage (\%) & Bias (\%) & Coverage (\%) \\
\midrule
Baseline & 0.1 & 93 & $-1.3$ to $-0.4$ & 95.5 to 96 \\
Dimension 256, common shocks doubled & $-0.4$ & 93.5 & $-1.1$ to $-0.6$ & 94.5 to 96.5 \\
Text share 1.6\% & $-1.5$ & 93.5 & $-3.8$ to $-2.0$ & 92.5 to 94 \\
Text share 0.4\% & $-4.7$ & 90 & $-8.0$ to $-3.4$ & 93.5 to 95 \\
Nonlinear content, projected concept & $-0.1$ & 90 & $-0.6$ to $-0.2$ & 93.5 to 94 \\
Nonlinear content, full concept & $-4.4$ & 5 & 13 to 26 & 28.5 to 55 \\
\bottomrule
\end{tabular}}
\floatnote{The table reports the bias and the coverage of the estimator in simulated panels. Bias is the mean estimate less the true value, in percent of the true value. Coverage is the frequency with which the nominal 95\% confidence interval contains the true value. Ranges are over the three horizons. In the last two rows the same estimates are compared with two different true values.}
\end{table}


\begin{thebibliography}{99}
\bibitem[Alain and Bengio(2017)Alain and Bengio]{alain2017understanding}
Alain, G., and Y. Bengio. 2017. ``Understanding Intermediate Layers Using Linear Classifier Probes.'' Workshop Paper, International Conference on Learning Representations, arXiv:1610.01644.

\bibitem[Amihud(2002)Amihud]{amihud2002illiquidity}
Amihud, Y. 2002. Illiquidity and Stock Returns: Cross-Section and Time-Series Effects. \emph{Journal of Financial Markets} 5: 31--56.

\bibitem[Bai et~al.(2016)Bai, Philippon, and Savov]{bai2016have}
Bai, J., T. Philippon, and A. Savov. 2016. Have Financial Markets Become More Informative? \emph{Journal of Financial Economics} 122: 625--654.

\bibitem[Battaglia et~al.(2024)Battaglia, Christensen, Hansen, and Sacher]{battaglia2024inference}
Battaglia, L., T. Christensen, S. Hansen, and S. Sacher. 2024. ``Inference for Regression with Variables Generated by {AI} or Machine Learning.'' Working Paper, arXiv:2402.15585.

\bibitem[Belinkov(2022)Belinkov]{belinkov2022probing}
Belinkov, Y. 2022. Probing Classifiers: Promises, Shortcomings, and Advances. \emph{Computational Linguistics} 48: 207--219.

\bibitem[Ben-David and Chinco(2026)Ben-David and Chinco]{bendavid2026crimes}
Ben-David, I., and A. Chinco. 2026. ``Crimes Against Campbell-Shiller.'' Working Paper 35189, National Bureau of Economic Research.

\bibitem[Bernard and Thomas(1989)Bernard and Thomas]{bernard1989post}
Bernard, V. L., and J. K. Thomas. 1989. Post-Earnings-Announcement Drift: Delayed Price Response or Risk Premium? \emph{Journal of Accounting Research} 27: 1--36.

\bibitem[Boudoukh et~al.(2019)Boudoukh, Feldman, Kogan, and Richardson]{boudoukh2019information}
Boudoukh, J., R. Feldman, S. Kogan, and M. Richardson. 2019. Information, Trading, and Volatility: Evidence from Firm-Specific News. \emph{Review of Financial Studies} 32: 992--1033.

\bibitem[Brogaard et~al.(2022)Brogaard, Nguyen, Putnins, and Wu]{brogaard2022moves}
Brogaard, J., T. H. Nguyen, T. J. Putnins, and E. Wu. 2022. What Moves Stock Prices? The Roles of News, Noise, and Information. \emph{Review of Financial Studies} 35: 4341--4386.

\bibitem[Bybee et~al.(2024)Bybee, Kelly, Manela, and Xiu]{bybee2024business}
Bybee, L., B. Kelly, A. Manela, and D. Xiu. 2024. Business News and Business Cycles. \emph{Journal of Finance} 79: 3105--3147.

\bibitem[Bybee et~al.(2023)Bybee, Kelly, and Su]{bybee2023narrative}
Bybee, L., B. Kelly, and Y. Su. 2023. Narrative Asset Pricing: Interpretable Systematic Risk Factors from News Text. \emph{Review of Financial Studies} 36: 4759--4787.

\bibitem[Campbell(1991)Campbell]{campbell1991variance}
Campbell, J. Y. 1991. A Variance Decomposition for Stock Returns. \emph{Economic Journal} 101: 157--179.

\bibitem[Campbell and Shiller(1988)Campbell and Shiller]{campbell1988dividend}
Campbell, J. Y., and R. J. Shiller. 1988. The Dividend-Price Ratio and Expectations of Future Dividends and Discount Factors. \emph{Review of Financial Studies} 1: 195--228.

\bibitem[Chan(2003)Chan]{chan2003stock}
Chan, W. S. 2003. Stock Price Reaction to News and No-News: Drift and Reversal After Headlines. \emph{Journal of Financial Economics} 70: 223--260.

\bibitem[Chen et~al.(2013)Chen, Da, and Zhao]{chen2013drives}
Chen, L., Z. Da, and X. Zhao. 2013. What Drives Stock Price Movements? \emph{Review of Financial Studies} 26: 841--876.

\bibitem[Chen et~al.(2022)Chen, Kelly, and Xiu]{chen2022expected}
Chen, Y., B. T. Kelly, and D. Xiu. 2022. ``Expected Returns and Large Language Models.'' Working Paper, SSRN 4416687.

\bibitem[Chernozhukov et~al.(2018)Chernozhukov, Chetverikov, Demirer, Duflo, Hansen, Newey, and Robins]{chernozhukov2018double}
Chernozhukov, V., D. Chetverikov, M. Demirer, E. Duflo, C. Hansen, W. Newey, and J. Robins. 2018. Double/Debiased Machine Learning for Treatment and Structural Parameters. \emph{Econometrics Journal} 21: C1--C68.

\bibitem[Cho et~al.(2024)Cho, Kremens, Lee, and Polk]{cho2024scale}
Cho, T., L. Kremens, D. Lee, and C. Polk. 2024. Scale or Yield? {A} Present-Value Identity. \emph{Review of Financial Studies} 37: 950--988.

\bibitem[Collins et~al.(1994)Collins, Kothari, Shanken, and Sloan]{collins1994lack}
Collins, D. W., S. P. Kothari, J. Shanken, and R. G. Sloan. 1994. Lack of Timeliness and Noise as Explanations for the Low Contemporaneous Return-Earnings Association. \emph{Journal of Accounting and Economics} 18: 289--324.

\bibitem[Costello et~al.(2024)Costello, Levy, and Nikolaev]{costello2024uncovering}
Costello, A., B. Levy, and V. Nikolaev. 2024. ``Uncovering Information: Can {AI} Tell Us Where to Look?'' Working Paper.

\bibitem[Didisheim et~al.(2026)Didisheim, Kelly, Pourmohammadi, and Tian]{didisheim2026inefficient}
Didisheim, A., B. Kelly, M. Pourmohammadi, and H. Tian. 2026. ``The Inefficient Pricing of News.'' Working Paper 35093, National Bureau of Economic Research.

\bibitem[Dong et~al.(2024)Dong, Fan, and Peng]{dong2024fnspid}
Dong, Z., X. Fan, and Z. Peng. 2024. ``{FNSPID}: A Comprehensive Financial News Dataset in Time Series.'' In \emph{Proceedings of the 30th ACM SIGKDD Conference on Knowledge Discovery and Data Mining}, 4918--4927. Association for Computing Machinery.

\bibitem[Driscoll and Kraay(1998)Driscoll and Kraay]{driscoll1998consistent}
Driscoll, J. C., and A. C. Kraay. 1998. Consistent Covariance Matrix Estimation with Spatially Dependent Panel Data. \emph{Review of Economics and Statistics} 80: 549--560.

\bibitem[Fieller(1954)Fieller]{fieller1954some}
Fieller, E. C. 1954. Some Problems in Interval Estimation. \emph{Journal of the Royal Statistical Society: Series B} 16: 175--185.

\bibitem[Gentzkow et~al.(2019)Gentzkow, Kelly, and Taddy]{gentzkow2019text}
Gentzkow, M., B. Kelly, and M. Taddy. 2019. Text as Data. \emph{Journal of Economic Literature} 57: 535--574.

\bibitem[He et~al.(2025)He, Lv, Manela, and Wu]{he2025chronologically}
He, S., L. Lv, A. Manela, and J. Wu. 2025. ``Chronologically Consistent Large Language Models.'' Working Paper, arXiv:2502.21206.

\bibitem[Ke et~al.(2019)Ke, Kelly, and Xiu]{ke2019predicting}
Ke, Z. T., B. T. Kelly, and D. Xiu. 2019. ``Predicting Returns with Text Data.'' Working Paper 26186, National Bureau of Economic Research.

\bibitem[Kelly et~al.(2026)Kelly, Malamud, Schwab, and Xu]{kelly2026scaling}
Kelly, B. T., S. Malamud, J. Schwab, and T. A. Xu. 2026. ``Scaling Point-in-Time Language Models.'' Working Paper 35247, National Bureau of Economic Research.

\bibitem[Knox and Vissing-Jorgensen(2022)Knox and Vissing-Jorgensen]{knox2022decomposition}
Knox, B., and A. Vissing-Jorgensen. 2022. ``A Stock Return Decomposition Using Observables.'' Finance and Economics Discussion Series 2022-014, Board of Governors of the Federal Reserve System.

\bibitem[Koh et~al.(2020)Koh, Nguyen, Tang, Mussmann, Pierson, Kim, and Liang]{koh2020concept}
Koh, P. W., T. Nguyen, Y. S. Tang, S. Mussmann, E. Pierson, B. Kim, and P. Liang. 2020. ``Concept Bottleneck Models.'' In \emph{Proceedings of the 37th International Conference on Machine Learning}, 5338--5348. PMLR.

\bibitem[Kothari and Sloan(1992)Kothari and Sloan]{kothari1992information}
Kothari, S. P., and R. G. Sloan. 1992. Information in Prices About Future Earnings: Implications for Earnings Response Coefficients. \emph{Journal of Accounting and Economics} 15: 143--171.

\bibitem[Lerman and Livnat(2010)Lerman and Livnat]{lerman2010new}
Lerman, A., and J. Livnat. 2010. The New {Form 8-K} Disclosures. \emph{Review of Accounting Studies} 15: 752--778.

\bibitem[Lochstoer and Tetlock(2020)Lochstoer and Tetlock]{lochstoer2020drives}
Lochstoer, L. A., and P. C. Tetlock. 2020. What Drives Anomaly Returns? \emph{Journal of Finance} 75: 1417--1455.

\bibitem[L{\'o}pez de Prado(2018)L{\'o}pez de Prado]{lopezdeprado2018advances}
L{\'o}pez de Prado, M. 2018. \emph{Advances in Financial Machine Learning}. Hoboken, NJ: Wiley.

\bibitem[Lopez-Lira and Tang(2023)Lopez-Lira and Tang]{lopezlira2023chatgpt}
Lopez-Lira, A., and Y. Tang. 2023. ``Can {ChatGPT} Forecast Stock Price Movements? Return Predictability and Large Language Models.'' Working Paper, arXiv:2304.07619.

\bibitem[Loughran and McDonald(2011)Loughran and McDonald]{loughran2011liability}
Loughran, T., and B. McDonald. 2011. When Is a Liability Not a Liability? Textual Analysis, Dictionaries, and 10-Ks. \emph{Journal of Finance} 66: 35--65.

\bibitem[Ludwig et~al.(2025)Ludwig, Mullainathan, and Rambachan]{ludwig2025large}
Ludwig, J., S. Mullainathan, and A. Rambachan. 2025. ``Large Language Models: An Applied Econometric Framework.'' Working Paper 33344, National Bureau of Economic Research.

\bibitem[Lundholm and Myers(2002)Lundholm and Myers]{lundholm2002bringing}
Lundholm, R., and L. A. Myers. 2002. Bringing the Future Forward: The Effect of Disclosure on the Returns-Earnings Relation. \emph{Journal of Accounting Research} 40: 809--839.

\bibitem[Meursault et~al.(2023)Meursault, Liang, Routledge, and Scanlon]{meursault2023pead}
Meursault, V., P. J. Liang, B. R. Routledge, and M. M. Scanlon. 2023. {PEAD.txt}: Post-Earnings-Announcement Drift Using Text. \emph{Journal of Financial and Quantitative Analysis} 58: 2299--2326.

\bibitem[Roll(1988)Roll]{roll1988r2}
Roll, R. 1988. {$R^2$}. \emph{Journal of Finance} 43: 541--566.

\bibitem[Sarkar and Vafa(2024)Sarkar and Vafa]{sarkar2024lookahead}
Sarkar, S. K., and K. Vafa. 2024. ``Lookahead Bias in Pretrained Language Models.'' Working Paper, SSRN 4754678.

\bibitem[Savor(2012)Savor]{savor2012stock}
Savor, P. G. 2012. Stock Returns After Major Price Shocks: The Impact of Information. \emph{Journal of Financial Economics} 106: 635--659.

\bibitem[Siano(2025)Siano]{siano2025news}
Siano, F. 2025. The News in Earnings Announcement Disclosures: Capturing Word Context Using {LLM} Methods. \emph{Management Science} 71: 9831--9855.

\bibitem[Tetlock(2007)Tetlock]{tetlock2007giving}
Tetlock, P. C. 2007. Giving Content to Investor Sentiment: The Role of Media in the Stock Market. \emph{Journal of Finance} 62: 1139--1168.

\bibitem[Tetlock(2010)Tetlock]{tetlock2010does}
Tetlock, P. C. 2010. Does Public Financial News Resolve Asymmetric Information? \emph{Review of Financial Studies} 23: 3520--3557.

\bibitem[Thompson(2011)Thompson]{thompson2011simple}
Thompson, S. B. 2011. Simple Formulas for Standard Errors That Cluster by Both Firm and Time. \emph{Journal of Financial Economics} 99: 1--10.

\bibitem[Vuolteenaho(2002)Vuolteenaho]{vuolteenaho2002drives}
Vuolteenaho, T. 2002. What Drives Firm-Level Stock Returns? \emph{Journal of Finance} 57: 233--264.

\bibitem[Williamson et~al.(2021)Williamson, Gilbert, Carone, and Simon]{williamson2021nonparametric}
Williamson, B. D., P. B. Gilbert, M. Carone, and N. Simon. 2021. Nonparametric Variable Importance Assessment Using Machine Learning Techniques. \emph{Biometrics} 77: 9--22.

\bibitem[Wu et~al.(2025)Wu, Akin, Martineau, Gr{\'e}goire, and Veneris]{wu2025extracting}
Wu, Y., E. M. Akin, C. Martineau, V. Gr{\'e}goire, and A. Veneris. 2025. ``Extracting the Structure of Press Releases for Predicting Earnings Announcement Returns.'' Working Paper, arXiv:2509.24254.
\end{thebibliography}
\end{document}